\documentstyle[12pt]{article}
\begin{document}
\hoffset=-1truecm
\voffset=-2truecm
\begin{center}
{\bf Integrable Heisenberg- van Vleck chains\\
 with variable range exchange}\\
V.I.Inozemtsev\\
{\small YITP,
Kyoto University, Kyoto 606-8502, Japan\\
and\\
BLTP JINR,
141980 Dubna, Moscow Region, Russia}\footnote{Permanent address}\\

Abstract\\
\end{center}
{\small The review of recent results in the s=1/2 quantum spin chains
with $1/\sinh^2(\kappa r)$ exchange is presented. Related problems in the theory
of classical and quantum Calogero-Sutherland-Moser systems with inverse square
hyperbolic and elliptic potentials are discussed. The attention is paid
to finding the explicit form of corresponding  Bethe-Ansatz equations
and to connection with generalized Hubbard chains in one dimension.} \\
 
\noindent {\bf 1. Introduction}\\
The idea of spin exchange interaction of electrons as natural explanation
of ferromagnetism was first proposed by Heisenberg [1] and soon realized 
in mathematical form by Dirac [2]. But the first appearance of the famous
Heisenberg Hamiltonian in solid-state physics occured three years later
in the book by van Vleck [3]. Now it is of common use and was investigated 
from many points of view by various methods of condensed matter theory.
In two and higher dimensions, the problem of finding the eigenvalues and
eigenvectors can be solved only by approximate or numerical methods. In one
dimension the {\it exact} solution was obtained in the seminal paper by Bethe [4]
who considered most important case of nearest-neighbor exchange 
described by the Hamiltonian
$$H= \sum_{1\leq{j}\neq{k}\leq{N}}h(j-k)(\vec\sigma_{j}\vec\sigma_{k}-1),
\eqno(1)$$
where $\{\sigma_{j}\}$ are the usual Pauli  matrices acting on the s=1/2 spin
located at the site $j$ and {\it exchange constants} $\{h\}$ are of extreme
short-range form,
$$h(j)=J(\delta_{\vert j\vert ,1}+\delta_{\vert j\vert ,N-1}). \eqno(2)$$
It turned out that the solution comes in the form of linear combinations of plane
waves chosen as to satisfy certain conditions required by (1,2).

Starting with this solution known as {\it Bethe Ansatz}, the investigation
of one-dimensional exactly solvable models of interacting objects
(spins, classical and quantum particles) has given a number of results both
of physical and mathematical significance [5]. Bethe found his solution
empirically; at that time the possibility to solve the quantum-mechanical
problems was not associated with the existence of underlying symmetry. The role
of such symmetries has been recognized much later, with one of the highlights
being the Yang-Baxter equation which allows one to find some regular way to
finding new examples of exactly solvable models [6,7]. In many cases, however,
the empirical ways are more productive since they use some physical information
on their background and are not so complicated from mathematical viewpoint.

This concerns especially to the Calogero-Sutherland-Moser (CSM) models which were
discovered about thirty years ago. They describe the motion of an
arbitrary number of classical and quantum nonrelativistic particles interacting
via two-body singular potentials with the Hamiltonian
$$ H_{CSM}=\sum_{j=1}^{M}{{p_{j}^{2}}\over 2} +l(l+1)\sum_{j<k}^{M}V(x_{j}
-x_{k}), \eqno(3)$$
where  $\{p,q\}$ are canonically conjugated momenta and positions of particles, 
$l \in {\bf R}$ and the two-body potentials are of the form:
$$V(x)={1\over {x^2}}, \qquad {{\kappa^2}\over {\sin^2(\kappa x)}}, \eqno (4)$$
$$V(x)={{\kappa^2}\over {\sinh^2(\kappa x)}},\eqno (5)$$
$$V(x)=\wp(x),\eqno(6)$$
where $\kappa \in {\bf R}_{+}$ and $\wp(x)$ is the double periodic Weierstrass $\wp$ function
determined by its two periods $\omega_{1}\in {\bf R}_{+}$, $\omega_{2}=i\pi/\kappa$ as
$$\wp(x)={1\over {x^2}}+\sum_{m,n\in {\bf Z}, m^{2}+n^{2}
\neq 0}\left[{1\over {(x-m\omega_{1}-n\omega_{2})^2}}
-{1\over {(m\omega_{1}+n\omega_{2})^2}}\right].\eqno(7)$$
The solvability of the eigenproblem for first two potentials (4) has been found independently
 by Calogero [8] and Sutherland [9] in quantum case while (5) and (6) have been found much later
[10-11] for classical particles by constructing extra integrals of motion (conserved quantities)
via the method of Lax pair. Namely, it turned out that the dynamical equations of motion
are equivalent to the $(M\times M)$ matrix relation
$${{dL}\over {dt}}=[L,M], \eqno (8)$$
where
$$L_{jk}=p_{j}\delta_{jk}+(1-\delta_{jk})f(x_{j}-x_{k}), \eqno(9)$$
$$M_{jk}=(1-\delta_{jk})g(x_{j}-x_{k})-\delta_{jk}\sum_{m\neq j}^{M}V(x_{j}-x_{m}), 
$$
if the functions $f,g,V$ obey the Calogero-Moser functional equation
$$ f(x)g(y)-f(y)g(x)=f(x+y)[V(y)-V(x)]\eqno(10)$$
which implies $g(x)=f'(x)$, $V(x)=-f(x)f(-x)$. The most general form of the solution to (10)
has been found by Krichever [12] in terms of the Weierstrass sigma functions which give rise to the potential (6).
Note that (4) might be considered as limits of (5,6) as $\kappa\to 0$,
and (6) can be regarded as double periodic form of (5) ((5) under periodic boundary conditions).
The existence of $M$ functionally independent integrals of motion in 
involution  follows from the evident relations $d({\rm tr}L^n )/dt=0$, $1\leq n \leq M$.
The fact that all these conserved quantities are in involution also follows from functional
equation (10) but needs some cumbersome calculations which can be extended also to the quantum case
where the time derivative should be replaced to quantum commutator with Hamiltonian [13].
In the review paper [13], one can find a lot of interesting facts
about the quantum models (3-6) established till 1983.

The Bethe Ansatz technique and the theory of CSM models developed independently till
1988 when Haldane [14] and Shastry [15] proposed a new spin 1/2 model with long-range
exchange resembling (4),
$$  h(j)=J\left({\pi\over N}\right)^2 \sin ^{-2}\left({{\pi j}\over N}\right), \eqno(11)$$
which has very simple ground-state function of Jastrow type in the antiferromagnetic regime
J$>$0 and many degeneracies in the full spectrum. The complete integrabilty of the model
and the reason of these degeneracies-the $sl(2)$ Yangian symmetry has been understood later-
for a comprehensive review see [16] and references therein. The Haldane-Shastry model has
many nice features, including the interpretation of the excited states as ideal "spinon" gas,
exact calculation of the partition function in the thermodynamic limit and
the possibility of exact calculation of various correlations 
in the antiferromagnetic ground state [16]. 

The connection with the Bethe case of 
nearest-neighbor exchange also came soon: in 1989, I have found that the Bethe and 
Haldane-Shastry forms of exchange are in fact the limits of more general model in which $h(j)$
is given by the elliptic Weierstrass function in complete analogy with (6),
$$h(j)=J\wp_{N}(j), \eqno (12)$$
where the notation $\wp_{N}$ means that the real period of the Weierstrass function equals $N$.
The absolute value of the second period $\pi/\kappa$ is a free parameter of the model [17].
The Haldane-Shastry spin chain arises as a limit of $\kappa\to 0$.
When considering the case of an infinite lattice ($N\to \infty$), one recovers the hyperbolic
form of exchange (5) which degenerates into the nearest-neighbor exchange if $\kappa \to
\infty$ under proper normalization of the coupling constant $J$: $J\to \sinh^2(\kappa)J$.
Hence (6) might be regarded as (5) under periodic boundary conditions (finite lattice).
Various properties of hyperbolic and elliptic spin chains form the main subject of the present
review.

The analogy between quantum spin chains and CSM models is much deeper than simple similarity
of spin exchange constants and two-body CSM potentials. It concerns mainly in similarities
in the form of {\it wave functions} of discrete and continuous cases. Namely, already in [17]
it has been mentioned that the solution of two-magnon problem for the exchange (12) and its
degenerated hyperbolic form can be obtained via two-body CSM systems with potentials (5,6)
at $l=1$;  it has also been found soon to be true for three- and four-magnon wave functions for
hyperbolic exchange [18]. Why does this similarity hold? Till now this is poorely understood,
but it is working even for elliptic case as it will be shown in Sections {\bf 3-4}. Another
question concerns integrability of the spin chains with hyperbolic and elliptic exchange, i.e.
the existence of a family of operators commuting with the Hamiltonian. In the case of
nearest-neighbor exchange, such a family can be easily found within the framework of the
quantum inverse scattering method [6]. However, it is not clear up to now how this method
should be used in the hyperbolic and (more general) elliptic cases. Instead, in Section {\bf 2}
the Lax pair and empirical way of constructing conserved quantities is exposed. Section {\bf 5}
contains various results for hyperbolic models on inhomogeneous lattices defined as 
equlibrium positions of the {\it classical} CSM hyperbolic systems in various external fields. 
Recent results concerning the integrability of the related Hubbard chains with variable range 
hopping are presented in Section {\bf 6}. The list of still unsolved problems is given in the
last Section {\bf 7} which contains also a short summary and discussion.\\

\noindent {\bf 2. Lax pair and integrability}\\

\noindent I shall consider in this Section a bit more general models with the Hamiltonian
$${\cal H}_{N}={1\over 2}\sum_{1\leq j\neq k\leq N}h_{jk}P_{jk}, \eqno(13)$$
where $\{P_{jk}\}$ are operators of an arbitrary representation of the permutation group
$S_{N}$. The spin chains discussed above fall into this class of models, as it follows
from the spin representation of the permutation group:
$$P_{jk}={1\over 2}(1+\vec\sigma_{j}\vec\sigma_{k}).$$ 
The {\it exchange constants} $h_{jk}$ in (13) are supposed to be translation invariant.
The notation $\psi_{jk}=\psi (j-k)$ will be assumed for any function of the difference
of numbers $j$ and $k$ in this Section. The problem is: how to select the function $h$
so as to get a model with integrals of motion commuting with the Hamiltonian (13)? 
The answer has been done in [17]: one can try to construct for the model the {\it
quantum} Lax pair analogous to (9,10) with $N\times N$ matrices:
$$L_{jk}=(1-\delta_{jk})f_{jk}P_{jk}, \quad M_{jk}=(1-\delta_{jk})g_{jk}P_{jk}-\delta_{jk}
\sum_{s\neq j}^{N}h_{js}P_{js}.$$
The quantum Lax relation $[{\cal H},L]=[L,M]$ is equivalent to functional
Calogero-Moser equation for $f,g,h$: 
$$
f_{pq}g_{qr}-g_{pq}f_{qr}=f_{pr}(h_{qr}-h_{pq})  \eqno(14)
$$
supplemented by the periodicity condition
$$
h'_{pq}=h'_{p,q+N}, \eqno(15)
$$
where $h'(x)$ is an odd function of its argument,
$
h'_{pq}=f_{qp}g_{pq}-f_{pq}g_{qp}.
$
The most general solution to (14) has been given in [12] as the combination
of the Weierstrass sigma functions. There is the normalization of $f$ and $h$
which allows one to write the relations
$$
h(x)=f(x)f(-x),\quad g(x)={{df(x)}\over{dx}},\quad h'(x)={{dh(x)}
\over{dx}}.    
$$
The solution given in [12] looks as
$$
f(x)={{\sigma(x+\alpha)}\over{\sigma(x)\sigma(\alpha)}}
\exp(-x\zeta(\alpha)),\qquad h(x)=\wp(\alpha)-\wp(x).  \eqno(16)
$$
where 
$$\zeta'(x)=-\wp(x),\quad {{d (\log\sigma(x))}\over {dx}}=\zeta (x)$$
and $\alpha$ is the spectral parameter which does not introduce any new
in exchange dynamics. The periodicity condition (15) means that all Weierstrass
functions in (16) are defined on the torus
${\bf T}_{N}={\bf C}/N{\bf Z}+i{\pi\over\kappa}{\bf Z}$, $\kappa\in {\bf R}_{+}$ is the free parameter
of the model. It is easy to see that the exchange (16) reduces in the limit
$\kappa\to\ 0$ to the Haldane-Shastry model and the limit of infinite lattice
size $(N\to \infty)$ corresponds to the hyperbolic variable range form of exchange.
And finally, in the limit $\kappa\to \infty$ just nearest-neighbor exchange (2) is reproduced as it was already mentioned in the preceding Section. 

However, the problem is not classical one and the existence of the Lax 
representation does not guarantee  the existence of the integrals of motion as
invariants of the $L$ matrix. In fact, just for the problem under consideration
the operators ${\rm tr}L^{n}$ do {\it not} commute with the Hamiltonian.
Nevertheless, already in [17] the way of constructing integrals of motion
on the base of $f$ function of Lax pair was proposed. Namely, it was
found that the operator
$$ J(\alpha)=\sum_{j\neq k\neq l}f_{jk}f_{kl}f_{lj}P_{jk}P_{kl}\eqno (17)$$
commutes with ${\cal H}$! Moreover, the dependence of the right-hand side
on the spectral parameter implies that there are two functionally independent
operators bilinear in $\{P\}$ commuting with ${\cal H}$,
$$J_{1}=\sum_{j\neq k\neq  l}(\zeta(j-k)+\zeta(k-l)+\zeta(l-j))P_{jk}P_{kl},$$
$$J_{2}=\sum_{j\neq k \neq l}[2(\zeta(j-k)+\zeta(k-l)+\zeta(l-j))^3+$$
$$\wp'(j-k)+\wp'(k-l) +\wp'(l-j)]P_{jk}P_{kl}.$$
Very long but straightforward calculations show that $J_{1,2}$ mutually commute.

It turns out [32] that the construction (17) can be generalized for more complicated
operators with higher degrees of $\{P\}$. The basic idea is to use the operators
of cyclic permutations $P_{s_{1}...s_{l}}\equiv P_{s_{1}s_{2}}P_{s_{2}s_{3}}...
P_{s_{l-1}s_{l}}$ and functions
$F_{s_{1}..s_{l}}=f_{s_{1}s_{2}}f_{s_{2}s_{3}}..$ 
$f_{s_{l-1}s_{l}}f_{s_{l}s_{1}}$ which are invariant under the action of elements
of a group of cyclic permutations of subindices $(1,..l)$. If one denotes as
$\Phi(s_{1},..,s_{l})$ the functions which are completely symmetric in their
arguments, and 
$\sum_{C\in{C}_{l}}B_{s_{1}..s_{l}}$ as the sum over all cyclic permutations of the
subindices of $B_{s_{1}..s_{l}}$, the following properties of the above objects
are useful:\\
(A) {\it The functions $h(x)$ and $h'(x)$ obey the relation}
$$
\sum_{C\in{C}_{3}}h'_{s_{1}s_{2}}(h_{s_{1}s_{3}}-h_{s_{2}s_{3}})=0.
$$
(B) {\it The sum $F_{s_{1}..s_{l+1}}^{(C)}=
\sum_{C\in{C}_{l}}F_{s_{1}..s_{l}s_{l+1}}$ does not
depend on} $s_{l+1}$.\\
(C) {\it The sum $\sum_{s_{1}\neq..s_{l+1}}\Phi(s_{1},..,s_{l+1})
F_{s_{1}..s_{l}}(h_{s_{l+1}s_{l}}-h_{s_{l+1}s_{1}})P_{s_{1}..s_{l+1}}$
vanishes for any symmetric function} $\Phi$.\\
(D) {\it The sum
$$
S_{l}(\Phi)=\sum_{s_{1}\neq..s_{l}}F_{s_{1}..s_{l}}\Phi(s_{1},..,s_{l})
(h_{s_{l}s_{l-1}}-h_{s_{l}s_{1}})P_{s_{1}...s_{l-1}}
$$
has a representation in the form $S_{l-1}^{(1)}(\Phi)+S_{l-1}^{(2)}(\Phi)$,
where
$$
S_{l-1}^{(1)}(\Phi)=
\sum_{s_{1}\neq..s_{l-1}}F_{s_{1}..s_{l-1}}\left(
{1\over{l-1}}\sum_{p\neq{s}_{1},..s_{l-1}}^{N}\Phi(s_{1},..,s_{l-1},p)
\sum_{\nu=1}^{l-1}h'_{ps_{\nu}}
\right)P_{s_{1}..s_{l-1}} 
$$
and
$$
S_{l-1}^{(2)}(\Phi)=
\sum_{s_{1}\neq..s_{l-1}}F_{s_{1}..s_{l-2}}(h_{s_{l-1}s_{l-2}}-
h_{s_{l-1}s_{1}})\left(
\sum_{p\neq{s}_{1},..s_{l-1}}h_{s_{l-1}p}\Phi(s_{1},..,s_{l-1},p)
\right)
P_{s_{1}..s_{l-1}}
$$
if} $l>3$,
$$
S_{2}^{(2)}(\Phi)=-{1\over2}
\sum_{s_{1}\neq{s}_{2}}h'_{s_{1}s_{2}}
\sum_{p\neq{s}_{1},s_{2}}^{N}(h_{s_{1}p}-h_{s_{2}p})\Phi(s_{1},s_{2},p)
P_{s_{1}s_{2}}.
$$
The main statement concerning the integrals of motion for the Hamiltonian (13)
can be  proved without the use of the specific form (16) of the
solution to the Calogero-Moser equation. It can be formulated as follows:
Let $I_{m}$ $(3\leq{m}\leq{N})$ be the linear combinations
of the operators of cyclic permutations in ordered sequences of $N$ symbols,
$$
I_{m}=\sum_{l=0}^{[{m\over2}]-1}{{(-1)^l}\over{m-2l}}
\sum_{s_{1}\neq..s_{m-2l}}
\Phi^{(l)}(s_{1},..,s_{m-2l})F_{s_{1}..s_{m-2l}}
P_{s_{1}...s_{m-2l}}.
\eqno(18)
$$
Then they will give the integrals of motion as it follows from\\
{\bf Proposition 2.1}.  The operators $I_{m}$ commute with ${\cal H}_{N}$
given by (13) if  the functions $\Phi^{(l)}$ are determined by the recurrence
relation
$$ \Phi^{(0)}=1, \quad \Phi^{(l)}(s_{1},..,s_{m-2l})=
l^{-1}\sum_{1\leq{j}<k\leq{N};j,k\neq{s}_{1},..s_{m-2l}}h_{jk}\Phi^{(l-1)}
(s_{1},..,s_{m-2l},j,k)
\eqno(19)
$$
or, equivalently, are given by sums over $2l$ indices
$$
\Phi^{(l)}(s_{1},..,s_{m-2l})=(l!)^{-1}\sum_{1\leq{j}_{\alpha}<k_{\alpha}
\leq{N};\{j,k\}\neq {s}_{1},..s_{m-2l}}\lambda_{\{jk\}}\prod_{\alpha=1}
^{l}h_{j_{\alpha}k_{\alpha}},
\eqno(20)
$$
where $\lambda_{\{jk\}}$ equals 1 if the product $\prod_{\alpha\neq\beta}
^{l}(j_{\alpha}-j_{\beta})(k_{\alpha}-k_{\beta})(j_{\alpha}-k_{\beta})$
differs from zero and vanishes otherwise.

The rigorous proof of the statements A-D  can be found in
[32]. Here I give only sketch of the proof of Proposition 2.1. It is based on the calculation of the commutator
$$
J_{n}=\sum_{s_{1}\neq..s_{n}}[\Phi(s_{1},..,s_{n})
F_{s_{1}..s_{n}}P_{s_{1}..s_{n}},{\cal H}_{N}],
\eqno(21)
$$where $\Phi$ is symmetric in its variables. With the use of
invariance of $F_{s_{1}..s_{n}}$ and $P_{s_{1}..s_{n}}$ under cyclic
changes of summation variables it is easy to show that this commutator can
be written as
$$
J_{n}= n\left[
J_{n}^{(1)}+J_{n}^{(2)}+\sum_{\nu=2}^{[n/2]}\left(
1-\left({{n-1}\over2}-[{n\over2}]\right)\delta_{\nu,[n/2]}\right)
J_{n,\nu}^{(3)}\right],
$$
where
$$
J_{n}^{(1)}=\sum_{s_{1}\neq..s_{n+1}}\Phi(s_{1},..,s_{n})
F_{s_{1}..s_{n}}(h_{s_{n}s_{n+1}}-h_{s_{1}s_{n+1}})
P_{s_{1}..s_{n+1}},
$$
$$
J_{n}^{(2)}=\sum_{s_{1}\neq..s_{n}}\Phi(s_{1},..,s_{n})
F_{s_{1}..s_{n}}(h_{s_{n-1}s_{n}}-h_{s_{1}s_{n}})
P_{s_{1}..s_{n-1}},
$$
$$
J_{n,\nu}^{(3)}=\sum_{s_{1}\neq..s_{n}}\Phi(s_{1},..,s_{n})
F_{s_{1}..s_{n}}(h_{s_{\nu}s_{n}}-h_{s_{1}s_{\nu+1}})
P_{s_{1}..s_{\nu}}P_{s_{\nu+1}..s_{n}}.
$$
The third term can be transformed with the use of functional equation (14)  and 
cyclic symmetry of $P_{s_{1}...s_{\nu}}$ and $P_{s_{\nu+1}..s_{n}}$ to the
form
$$
J_{n,\nu}^{(3)}=\sum_{s_{1}\neq..s_{n}}\Phi(s_{1},..,s_{n})\left[
\nu^{-1}\varphi_{s_{\nu+1}..s_{n-1}s_{n}}
(F_{s_{1}..s_{\nu}s_{\nu+1}}^{(C)}-
F_{s_{1}..s_{\nu}s_{n}}^{(C)})
\right.
$$
$$\left.+F_{s_{\nu+1}..s_{n}}(\varphi_{s_{\nu+1}s_{1}..s_{\nu}}-
\varphi_{s_{1}..s_{\nu+1}})\right]
P_{s_{1}..s_{\nu}}P_{s_{\nu+1}..s_{n}}
, \eqno(22)$$
where 
$
\varphi_{s_{1}s_{2}..s_{l+1}}=f_{s_{1}s_{2}}f_{s_{2}s_{3}}..
f_{s_{l}s_{l+1}}g_{s_{l+1}s_{1}}.        
$

Now it is easy to see that the term in the first brackets in (22) disappears due to
statement (B) and the term in the second brackets vanishes due to the relation
$$
F_{s_{1}..s_{l}}(h_{s_{1}s_{l+1}}-h_{s_{l}s_{l+1}})=
\varphi_{s_{1}s_{2}..s_{l+1}}-\varphi_{s_{l+1}s_{1}..s_{l}}, 
$$
which allows to transform this term to the expression which vanishes upon
symmetrization in all cyclic changes of $s_{1},..s_{\nu}$. Hence the operator
(21) contains only cyclic permutations of rank $(n+1)$ and $(n-1)$. This fact
leads to the idea of recurrence construction of the operators (18) which would
commute with the Hamiltonian. It happens if the functions $\{\Phi\}$ obey the 
recurrence relation
$$
\sum_{p\neq{s}_{1},..s_{m-2l-1}}(h_{ps_{m-2l-1}}-h_{ps_{1}})
\Phi^{(l)}(s_{1},..,s_{m-2l-1},p)
$$
$$
=\Phi^{(l+1)}(s_{1},s_{2},..,s_{m-2l-2})-\Phi^{(l+1)}(s_{m-2l-1},
  s_{2},..,s_{m-2l-2}), \Phi^{(0)}=1,  
$$
which can be solved in the form (19) or (20).\\
The dependence of (18-20) on the spectral parameter $\alpha$ via the relations
(16) allows one to conclude that there are several integrals of motion at each
$m$. Namely, the analysis of this dependence shows that the operators (18)
can be written in the form
$$
I_{m}=w_{m}(\alpha)P_{m}+\sum_{\mu=1}^{m-2}w_{m-\mu}(\alpha)I_{m,\mu}
+I_{m,m},
$$
where $P_{m}$ commutes with all operators of elementary transpositions, 
$I_{m,\mu}$ are linear combinations of $\{P_{s_{1}..
s_{m-2l}}\}$ which do not depend on $\alpha$, and $w_{m-\mu}(\alpha)$
are linearly independent elliptic functions of the spectral parameter.
Nothing is known for the mutual commutativity of these operators except
the explicit result for $m=3$ mentioned above. Still there is no connection
with Yang-Baxter theory, i.e. the corresponding $R$-matrix and $L$-operators
are unknown. There is an excellent paper by K.Hasegawa [33] which states that
$R$-matrix for spinless elliptic quantum Calogero-Moser systems is Belavin's one,
also of elliptic type. However, it is not clear how to extend Hasegawa's method
to the spin case so as to reproduce the rich variety of the operators
(18).\\

\noindent {\bf 3. The infinite chain}\\

\noindent On the infinite line, the model is defined by the Hamiltonian
$$H=-{J\over 2}\sum_{j\neq k}{{\kappa^2}\over {\sinh^2 \kappa (j-k)}}
(\vec\sigma_{j}\vec\sigma_{k}-1)/2, \eqno(23)$$
where $j,k\in {\bf Z}$. At these conditions, only ferromagnetic case $J>0$ is
well defined. The spectrum to be found consists of excitations over ferromagnetic
ground state $\vert 0>$ with all spins up which has zero energy. The energy of
one spin wave is just given by Fourier transform of the exchange in (23),
$$\varepsilon (p)= J\left\{-{1\over 2}\wp_{1}({{ip}\over{2\kappa}})
+{1\over2}\left[{p\over \pi}\zeta_{1}\left(
{{i\pi}\over {2\kappa}}\right)-\zeta_{1}\left({{ip}\over{2\kappa}}\right)\right]^2
-{{2i\kappa}\over \pi}\zeta_{1}\left({{i\pi}\over{2\kappa}}\right)\right\}, \eqno(24)$$
where Weierstass functions $\wp_{1}$, $\zeta_{1}$ are defined on the torus
${\bf T}_{1}={\bf C}/({\bf Z}+{{i\pi}\over \kappa}{\bf Z})$, i.e. $\wp_{1}$ has
the periods $(1, \omega =i\pi/\kappa)$.

{\bf 3.1. Two-magnon scattering}.
The two-magnon problem for the model (23) is already nontrivial. One has
to solve the difference equation for two-magnon wave function $\psi (n_{1},
n_{2})$ which is defined by the relation
$$\vert\psi>=\sum_{n_{1}\neq n_{2}}\psi(n_{1}, n_{2})s^{-}_{n_{1}}s^{-}_{n_{2}}
\vert 0>,$$
where the operator $\{s^{-}_{n_{\alpha}}\}$ reverses spin at the site 
$n_{\alpha}$ and $\vert\psi>$ is an eigenvector of the Hamiltonian (23). The solution
is based on the formula [17]
$$\sum_{k=-\infty}^{\infty}{{\kappa^2 \exp(ikp)}\over{\sinh^2[a(k+z)]}}
\coth \kappa(k+l+z)=-{{\sigma_{1}(z+r_{p})}\over {\sigma_{1}(z-r_{p})}}
\coth(\kappa l)\exp\left[{{pz}\over \pi}\zeta_{1}(\omega/2)\right]$$
$$\times \{\wp_{1}(z)-\wp_{1}(r_{p})+2\left[\zeta_{1}(r_{p})-{{2r_{p}}\over
\omega}\zeta\left({\omega\over 2}\right)+{\kappa\over {\sinh (2\kappa l)}}(1-
\exp(-ipl))\right]$$
$$\times (\zeta_{1}(z+r_{p})-\zeta_{1}(z)+2\zeta_{1}(r_{p})-\zeta_{1}(2r_{p}))\}
, \eqno(25)$$
where $r_{p}=-\omega p/4\pi$ and $l\in {\bf Z}$.

The proof of (25) is based on the quasiperiodicity of the sum on its left-hand side and the structure of its only singularity at the point $z=0$ on a torus ${\bf T}_{1}$ obtained by factorization of a complex plane on the lattice of periods
$(1,\omega)$. The structure of (25) allows one to show that the two-magnon
wave function is given by the formula
$$ \psi(n_{1}, n_{2})={{
e^{i(p_{1}n_{1}+p_{2}n_{2})}\sinh[\kappa(n_{1}-n_{2})+\gamma]+
e^{i(p_{1}n_{2}+p_{2}n_{1})}\sinh[\kappa(n_{1}-n_{2})-\gamma]}\over
{\sinh \kappa(n_{1}-n_{2})}},
\eqno(26)$$
the corresponding energy is
$$ \varepsilon^{(2)}(p_{1}, p_{2})=\varepsilon(p_{1})+\varepsilon(p_{2}),$$
where $\varepsilon (p_{i})$ are given by (24) and the phase $\gamma$ is
connected with pseudomomenta $p_{1,2}$ by the relation
$$\coth \gamma={1\over 2\kappa}\left[\zeta_{1}\left({{ip_{2}}\over{2\kappa}}\right)                                -\zeta_{1}\left({{ip_{1}}\over{2\kappa}}\right)
+{{p_{1}-p_{2}}\over \pi}\zeta_{1}\left({{i\pi}\over{2\kappa}}\right)\right]
.\eqno(27)$$
This gives, in the limit of $\kappa\to\infty$ $(\omega\to 0)$, just the expression
for the Bethe phase [4], and the additivity of magnon energies takes place.
The equation (27) can be rewritten in the form
$$\coth \gamma={{f(p_{1})-f(p_{2})}\over{2\kappa}},
$$
where
$$f(p)= {p\over \pi}\zeta_{1}\left({{i\pi}\over{2\kappa}}\right)-\zeta_{1}
\left({{ip}\over{2\kappa}}\right).
\eqno(28)$$
It admits also the representation
$$f(p)=i\kappa\cot {p\over 2}-\kappa\sum_{n=1}^{\infty}\left[\coth\left({{ip}\over 2}+\kappa n
\right)+\coth\left({{ip}\over 2}-\kappa n\right)\right].$$ 
If $p_{1,2}$ are real, the wave function (26) describes scattering of magnons.
The relatively simple form of (27) allows one to investigate the bound states
of two magnons in detail [34]. Namely, in these states the wave function must
vanish as
$\vert n_{1}-n_{2}\vert\to\infty$.
It means that $p_{1}$ and $p_{2}$ should be
complex with $P=p_{1}+p_{2}$ real. The simplest possibility is given by the
choice
$$p_{1}={P\over 2}+iq,\quad p_{2}={P\over 2}-iq,
$$
where $q$ is real, and one can always choose $q>0$ for convenience.
Then vanishing
of $\psi(n_{1},n_{2})$ as $\vert n_{1}-n_{2}\vert\to\infty$ is equivalent to
the condition
$$\coth\gamma(p_{1},p_{2})={{f(p_{1})-f(p_{2})}\over{2\kappa}}=1.
\eqno(29)
$$
The structure of the function (28) is crucial for the analysis. It is easy to 
see that it is odd and double quasiperiodic,
$$f(p)=-f(-p),\quad 
f(p+2\pi)=f(p),\quad
f(p+2i\kappa)=f(p)+2\kappa.
\eqno(30)$$
Note that one can always choose $q\leq \kappa$ due to (30). The equation (29)
can be rewritten in more detailed form
$$F_{P}(q)=1-{1\over 2\kappa}
\left[{{2iq}\over\pi}\zeta_{1}\left({{i\pi}\over{2\kappa}}\right)-
\zeta_{1}\left({{iP}\over{4a}}-{q\over{2\kappa}}\right)+
\zeta_{1}\left({{iP}\over{4a}}+{q\over{2\kappa}}\right)\right]=0.
\eqno(31)$$
At fixed real $P$ and $q$, the function (31) is real. Moreover, the relations
(30) imply the following properties of $F_{P}(q)$,
$$F_{P}(0)=1,\quad
F_{P}(q)=-F_{P}(2\kappa-q),\quad F_{P}(q)=-F_{P}(-q)+2.$$
One can immediately see that $F_{P}(\kappa)=0$ but this zero is unphysical: the
wave function in this point vanishes identically. The physical solution, 
if exists, must lie in the interval $0<q<\kappa$. Such a nontrivial zero exists if the derivative of $F_{P}(q)$ is positive at $q=\kappa$,  
$$F'_{P}(\kappa)=-{i\over {\pi \kappa}}\zeta_{1}\left({{i\pi}\over {2\kappa}}\right)
+{1\over {4\kappa^2}}\left[
\wp_{1}\left({{iP}\over{4\kappa}}-{1\over 2}\right)+
\wp_{1}\left({{iP}\over{4\kappa}}+{1\over 2}\right)\right]>0.
\eqno(32)$$
This inequality indeed takes place for the values of $P$ within the 
interval $0<P<P_{cr}$, $0<P_{cr}<\pi$ [34]. There should be at least 
one bound state specified by (31). At $P>P_{cr}$, the inequality
(32) does not hold and there are no bound states of this type (type I).

There is, however, another possibility for getting bound state. Since
$$f(p+i\kappa)=\kappa+i\chi(p), \quad f(p-i\kappa)=-\kappa +i\chi(p),$$
one gets only one real equation for real $\tilde p_{1,2}$ if one
puts $p_{1}=\tilde p_{1}+i\kappa,$ $p_{2}=\tilde p_{2}-i\kappa$,
 $$\chi(\tilde p_{1})-\chi(\tilde p_{2})=0.
\eqno(33)$$
Noting that
$$\chi(0)=\chi(\pi)=0,$$
$$  \chi'(0)={a\over 2}\sum_{n=-\infty}^{\infty}{1\over {\sinh^2 a\left(n+
{1\over 2}\right)}}>0,$$
$$\chi'\left({\pi\over 2}\right)
={a\over 2}\sum_{n=-\infty}^{\infty}{1\over {\sinh^2 \left(
{{i\pi}\over 4}+a\left(n+{1\over 2}\right)\right)}}<0,$$
it is easy to see that there should be 
 some value $\tilde p_{0}$ at
which $\chi(\tilde p)$ has a maximum on the interval $[0,\pi]$ and the
corresponding $\tilde p_{0}'=2\pi-\tilde p_{0}$ at which $\chi(\tilde p)$
has a minimum on the interval $[\pi,2\pi]$. As a matter of fact, $
\tilde p_{0}={{P_{cr}}\over 2}$. There are no other extrema of $\chi(\tilde p)$ on the interval $(0,\pi)$.
The presence of a maximum means that
the equation
$$\chi(\tilde p)=\chi_{0}$$
has two distinct real roots if $0\leq \chi_{0}<\chi\left({{P_{cr}}\over 2}
\right)$, ${{P_{cr}}\over 2}<\tilde p_{1}\leq \pi$ and $0\leq \tilde p_{2}
<{{P_{cr}}\over 2}$. These roots serve also as nontrivial solution to the
equation (33) and thus give the bound state of type II in which 
the wave function oscillates and decays exponentially as $\vert n_{1}-n_{2}
\vert\to\infty$. For $P_{cr}<P\leq \pi$ such a solution always exists.
Similar solutions corresponding to $-\chi\left({{P_{cr}}\over 2}\right)
< \chi_0 < 0$ can be found with any $-\pi\leq P <-P_{cr}$.

The above treatment is universal with respect to parameter $\kappa$ in the interval
$0<\kappa<\infty$. In the nearest-neighbor limit $\kappa\to\infty$, the type II states
with complex relative pseudomomentum and oscillating wave function disappear
$(P_{cr}\to\pi$ ) and the result coincides with
the known one for the Bethe solution.

{\bf 3.2. Multimagnon scattering}. After solving two-magnon problem, it
is natural to try to find a way to describing scattering of $M$ magnons
with $M\geq 3$, i.e. find solution to the difference equation
$$\sum_{\beta=1}^{M}\sum_{s \in{\bf Z}_{[n]}}V(n_{\beta}-s)\psi(n_{1},...n_{\beta-1},
s, n_{\beta+1},...n_{M})$$
$$=- \psi(n_{1},...n_{M})\left[\sum_{\beta\neq\gamma}^{M}V(n_{\beta}-n_{\gamma})
+J^{-1}\varepsilon_{M}-M\varepsilon_{0}\right],
\eqno(34)$$
where $n\in {\bf Z}^{M}$, the notation ${\bf Z}_{[n]}$ is used for the variety ${\bf Z}-(n_{1},...n_{M})$ and $\varepsilon_{0}=\sum_{j\neq 0}V(j)$. The 
exchange interaction $V(j)$ is of hyperbolic form (5).

The first attempt to solve (34) for $M>2$ was made in [18] with the use
of trial solution of the Bethe form and taking into account by semi-empirical
way the corrections needed due to non-local form of exchange in (34). In this
paper, the explicit solutions have been found for $M$=3,4 but the regular 
procedure of getting solution for higher values of $M$ was not proved 
rigorously. The rigorous treatment of the solutions to (34) has been found
later [21]. It is based on the analogy of the solution to (34) and 
corresponding solution to the quantum Calogero-Moser $M$-particle system with the same
two-body potential and specific value of the coupling constant in (3),
determined by $l=1$. This analogy is already seen in the form of two-magnon
wave function (26) and holds for $M=3,4$ too. It was the motivation 
of the paper [21] to use this analogy in detail.

The solution to $M$-particle system with hyperbolic potential and coupling constant with $l$=1 is not simple too. The first integral representation
for it has been obtained in [19] and more simple analytic form based on
recurrence operator relation was given in [20]. I will follow [20, 21] in
description of the $M$-magnon problem on an infinite lattice.

Let us start from continuum model (3) with the interaction (5) and $l$=1.
The solution can be written in the form
$$\chi^{(M)}_{p}(x)=\exp\left( i\sum_{\mu=1}^{M}ip_{\mu}x_{\mu}\right)\varphi
^{(M)}_{p}(x),
\eqno(35)
$$
where $\varphi^{(M)}_{p}(x) $ is periodic in each $x_{j}$,
$$\varphi^{(M)}_{p}(x)=\varphi_{p}^{(M)}( x_{1},...x_{j}+i\pi\kappa^{-1},...,x_{M}).$$
In [20], the explicit construction of the differential operator which
intertwines (3) at (5) and $l=1$ with the usual $M$- dimensional Laplasian has been proposed, and the functions 
of the type (35) have been represented in the form
$$\chi_{p}^{(M)}(x)=D_{M}\exp\left(i\sum_{\mu=1}^{M}p_{\mu}x_{\mu}\right),
\quad D_{M}=Q_{M}^{1...M-1}D_{M-1},
\eqno(36)$$
where
$$Q_{n}^{i_{1}...i_{m}}=Q_{n}^{i_{1}...i_{m-1}}\left[
{\partial\over{\partial x_{i_m}}}-{\partial\over{\partial x_{n}}}-2\kappa
\coth\kappa(x_{i_{m}}-x_{n})\right]$$
$$+\sum_{s=1}^{m-1}2\kappa^2 \sinh^{-2}[\kappa(x_{i_{s}}-x_{i_{m}})]
Q_{n}^{i_{1}...i_{s-1}i_{s+1}...i_{l-1}},\quad Q_{n}=1.
\eqno (37) $$
This double recurrence scheme is very cumbersome because of presence
of multiple differentiations but it allows one to reduce the construction 
of $\chi_{p}^{(M)}(x)$ to a much simple problem of solving the set
of linear equations. Indeed, it follows from (36) and (37) that the function
$\varphi_{p}^{M}(x)$ from (35) can be represented in the form
$$\varphi_{p}^{(M)}(x)=R(\{\coth\kappa(x_{j}-x_{k})\}),
\eqno(38)$$
where $R$ is a polynomial in the variables $\{\coth \kappa(x_{j}-x_{l})\}$.
As it can be seen from the structure of singularities in (3), the function
$\varphi_{p}^{M}(x)$ has a simple pole of the type $[\sinh\kappa(x_{j}-x_{k})
]^{-1}$ at each hyperplane $x_{j}-x_{k}=0$. As a consequence of (38), all the
limits of $\varphi_{p}^{(M)} (x)$ as $x_{j}\to\pm\infty$, must be finite. Combining
these properties with the periodicity of $\varphi$, one arrives at the following formula
for the eigenfunctions of the Calogero- Moser operator:
$$\chi_{p}^{(M)}(x)=\exp\left\{\sum_{\mu=1}^{M}[ip_{\mu}-\kappa(M-1)]x_{\mu}\right\}
\prod_{\mu>\nu}^{M}\sinh^{-1}\kappa(x_{\mu}-x_{\nu})S_{p}^{(M)}(y),
\eqno(39)$$
where $S_{p}^{(M)}(y)$ is a polynomial in $y_{\mu}=\exp(2\kappa x_{\mu})$
in which the maximal power of each variable cannot exceed $M-1$. Hence this
polynomial can be represented  in the form
$$S_{p}^{(M)}(y)=\sum_{m\in D^{M}}d_{m_{1}...m_{M}}(p)\prod_{\mu=1}^{M}y^{m_{\mu}},
\eqno (40)$$
where $D^M$ is the hypercube in ${\bf Z}^{M}$,
$$m\in D^{M}\leftrightarrow 0\leq m_{\beta}\leq M-1,$$
and $d_{m}(p) $ is the set of $M^M$ coefficients; it will be shown, however, that
most of them vanish. The eigenvalue condition for the function (39)
can be written in the form
$$\sum_{\beta=1}^{M}\left[2y_{\beta}{\partial\over {\partial y_{\beta}}}
\left(y_\beta{\partial \over {\partial y_{\beta}}}
+i\kappa^{-1}p_{\beta}-M+1\right)-i\kappa^{-1}p_{\beta}(M-1)+(M-1)(2M-1)/3\right]
S_{p}^{(M)}$$
$$-\sum_{\beta\neq\rho}^{M}{{y_{\beta}+y_{\rho}}\over {y_{\beta}-y_{\rho}}}
\left[y_{\beta}{\partial\over {\partial y_{\beta}}}
     -y_{\rho}{\partial\over {\partial y_{\rho}}}+{i\over{2\kappa}}
(p_{\beta}-p_{\rho})\right]S_{p}^{(M)}=0.
\eqno(41)$$
It can be satisfied if for each pair $(\beta, \rho)$ the polynomial 
$$\left[y_{\beta}{\partial\over{\partial y_{\beta}}}-y_{\rho}
                  {\partial\over {\partial y_{\rho}}}+{i\over{2\kappa}}
(p_{\beta}-p_{\rho})\right]S_{p}^{(M)}$$
is divisible by $(y_{\beta}-y_{\rho})$. With the use of (40) this condition gives 
$(M-1)(2M-1)M^{M}/2$ linear equations for the coefficients $d_{m}(p)$,
$$\sum_{n\in {\bf Z}}d_{m_{1}...m_{\beta}+n...m_{\rho}-n...m_{M}}(p)
\left[m_{\beta}-m_{\rho}+2n +{i\over {2\kappa}}(p_{\beta}-p_{\rho})\right]=0.
\eqno(42)$$ 
The sum over $n$ is finite due to restrictions to the indices of $d_{m}(p)$.
Substituting (40) gives also the set of equations
$$\sum_{m\in D^{M}}\left(\prod_{\mu=1}^{M}y_{\mu}^{m_{\mu}}\right)d_{m}(p)\{
\sum_{\beta=1}^{M}\left[2m_{\beta}^2 +{{2i}\over\kappa}p_{\beta}m_{\beta}
-\left(2m_{\beta}+{i\over\kappa}p_{\beta}-{{2M-1}\over 3}\right)(M-1)
\right]$$
$$ -\sum_{\beta\neq\rho}^{M}{{y_{\beta}+y_{\rho}}\over {y_{\beta}-y_{\rho}}}
\left[m_{\beta}-m_{\rho}+{i\over {2\kappa}}(p_{\beta}-p_{\rho})\right]\}
=0.
\eqno(43)$$
After performing explicit division by $(y_{\beta}-y_{\rho})$ in (43), one
gets finally the second system of $M^M$ equations. The structure of the set
$d_{m}(p)$ is specified by following propositions (for a sketch of proofs, see
[21]).

{\bf Proposition 3.1.} $S_{p}^{(M)}(y)$ is a homogeneous polynomial  of the 
degree $M(M-1)/2$.

 {\bf Proposition 3.2}. The set of $d_{m}(p)$ can be chosen as depending on 
$p$ and $\kappa$ only through combinations $\kappa^{-1}(p_{\mu}-p_{\nu})$.

{\bf Proposition 3.3.} Let $\{P\}$ be the following set of numbers $\{m_{\mu}\}$:
$m_{\mu}=P\mu-1$, where $P$ is an arbitrary permutation of the permutation group
$\pi_{M}$ and $1\leq \mu\leq M$. The nonvanishing $d_{m}(p)$ with coinciding values of $ \{m_{\mu}\}$ are expressed through $d_{\{P\}}(p)$. The latter
are determined by the system (42) up to some normalization constant $d_{0}$,
$$d_{\{P\}}(p)= d_{0}\prod_{\mu<\nu}^{M}\left[1+{i\over{2\kappa}}
(p_{P^{-1}\mu}-p_{P^{-1}\nu})\right].
\eqno(44)$$

{\bf Proposition 3.4.} Let $(-1)^P$ be the parity of the permutation $P$. 
If $x_{P(\mu +1)}-x_{P\mu}\to\+\infty$, $1\leq \mu\leq M-1$, then
$$\lim \chi_{p}^{(M)}(x)\exp\left(-i\sum_{\beta=1}p_{\beta}x_{\beta}\right)
=(-1)^P 2^{{{M(M-1)}\over 2}}d_{\{P^{-1}\}}(p).
\eqno(45)$$

According to Proposition 3.3, the solutions to (42) must obey (43), and (43)
has to be considered as a consequence of (42). Direct algebraic proof of this fact
is still absent. 

The problem is now to solve the equations (42). It can be done explicitly for
$M=3,4$ as follows: let $[\mu_{1}...\mu_{M}]$ be the permutation $(1\to \mu_{1},
...M\to \mu_{M})$ and $r_{\mu\nu}=i(2\kappa)^{-1}(p_{\mu}-p_{\nu}).$ Then, at
$M=3$ there are 6 coefficients of the $d_{\{P\}}$ type which are caculated by
the formula (44),
$$d_{012}(p)=d_{0}(1+r_{12})(1+r_{13})(1+r_{23}),
  d_{102}(p)=d_{0}(1+r_{21})(1+r_{23})(1+r_{13}), $$
$$d_{210}(p)=d_{0}(1+r_{32})(1+r_{31})(1+r_{21}),
  d_{021}(p)=d_{0}(1+r_{13})(1+r_{12})(1+r_{32}), $$
$$d_{120}(p)=d_{0}(1+r_{31})(1+r_{32})(1+r_{12}),
  d_{201}(p)=d_{0}(1+r_{23})(1+r_{21})(1+r_{31}).$$
The only nonvanishing coefficient of another type is determined from (42):
$$d_{111}(p)=d_{0}(6-r_{12}^2-r_{13}^2-r_{23}^2).$$
At $M=4$, there are 24 coefficients of $d_{\{P\}}$ type and other nonvanishing
terms with coinciding values of indices can be arranged in three sets. The first
two are given by elements with three coinciding indices and can be found
from (42) by using known expressions for $d_{\{P\}}$ type,
$$d_{1113}(p)=d_{0}(1+r_{14})(1+r_{24})(1+r_{34})
  (6-r_{12}^2-r_{13}^2-r_{23}^2),$$
$$d_{2220}(p)=d_{0}(1+r_{41})(1+r_{42})(1+r_{43})
  (6-r_{12}^2-r_{13}^2-r_{23}^2)$$
and other elements of these sets $d_{1131}(p),d_{1311}(p),d_{3111}(p)$ and
$d_{2202}(p), d_{2022}(p), d_{0222}(p)$ can be obtained by the permutations
[1243], [1342], [2341] of indices in these expressions. The remaining set
consists of the coefficients with two pairs of coinciding indices,
$$d_{1122}(p), d_{2211}(p), d_{2112}(p), d_{1221}(p), d_{1212}(p), d_{2121}(p).$$
They may be determined by (42) with the use of known coefficients belonging to
the first set,
$$ d_{1113}(p)(-2+r_{34})+d_{1122}(p)r_{34}+d_{1131}(p)(2+r_{34})=0.$$
and others come from the analogous equations arising after the permutations
[3412], [3214], [4123], [1324], [4123] of the indices.

These examples show that the solutions to (42) are crucial for determining
the whole function $\chi_{p}^{(M)}(x)$. The question is now to see how these
findings can be used for spin problem, i.e. the solution to the difference
equation (34). The motivation is the striking similarity of the wave functions
for $M=2$. Guiding by it, I proposed the multimagnon wave functions
similar to the functions like (35) with the structure (39), which are properly
symmetrized combinations of them,
$$\psi(n_{1},...n_{M})=\prod_{\mu\neq \nu}[\sinh\kappa(n_{\mu}-n_{\nu})]^{-1}
\sum_{P\in \pi_{M}}(-1)^{P}\exp\left(i\sum_{\lambda=1}^{M}p_{P\lambda}n_{\lambda}
\right)$$
$$\times \sum_{m\in D^{M}}\tilde d_{m_{1}...m_{n}}(p)\exp\left[\kappa\sum_{\lambda=1}^{M}(2m_{P\lambda}-M+1)
n_{\lambda}\right], 
\eqno(46)$$
where $\{\tilde d\}$ is the set of unknown coefficients which might be determined
from the $M$-magnon eigenequation if this Ansatz  is correct. To verify the
hypothesis (46), one has to calculate the left-hand side of the equation (34)
with wave function of the form (46),
$$L(\{n\})=\kappa^2 \sum_{\beta=1}^{M}\sum_{s\in {\bf Z}_{[n]}}[\sinh \kappa
(n_{\beta}-s)]^{-2}\psi(n_{1},...,n_{\beta-1},s, n_{\beta+1},...n_{M})$$
$$=\sum_{\beta=1}^{M}\sum_{P\in \pi_{M}}(-1)^{P}\left[\prod_{\mu>\nu;\mu,\nu
\neq \beta}^{M}\sinh\kappa(n_{\mu}-n_{\nu})\right]^{-1}(-1)^{\beta-1)}
\sum_{m\in D^{M}}\tilde d_{m_{1}...m_{n}}(p)$$
$$\times \exp\left\{\sum_{\gamma\neq \beta}[ip_{P\gamma}+\kappa
(2m_{P\gamma}-M+1)]n_{\gamma}\right\}W(p_{P\beta}, m_{P\beta}, \{n\}),
\eqno(47)
$$
where
$$W(p,m,\{n\})=\sum_{s\in {\bf Z}_{[n]}}{{\kappa^2}\over {\sinh^2\kappa
(s-n_{\beta})}}\prod_{\lambda\neq\beta}^{M}\sinh^{-1}\kappa (n_{\lambda}-s)$$
$$
\times\exp\{[ip+\kappa(2m-M+1)]s\}.
\eqno(48)$$
 The sum (48) converges for all $m\in D^{M}$ if $p\in {\bf C}$ is restricted to 
$\vert \Im m p\vert< 2\kappa$. The explicit calculation of the sum (48) is based
on the calculation of the function of a complex parameter $x\in {\bf C}$,
$$W_{q}(x)=\sum_{s\in {\bf Z}}{{\kappa^2\exp(qs)}\over {\sinh^{2}\kappa(s-n_{\beta}
+x)}}\prod_{\lambda\neq\beta}^{M}[\sinh\kappa(n_{\lambda}-s-x)]^{-1},
$$
$$q=ip+\kappa(2m-M+1).$$
As it follows from definition, this function is double quasiperiodic,
$$W_{q}(x+i\pi\kappa^{-1})=\exp[i\pi(M-1)]W_{q}(x),\quad W_{q}(x+1)=\exp (-q)
W_{q}(x).$$
Hence it can be treated on the torus ${\bf T}_{1}={\bf Z}/{\bf Z}+i\pi\kappa^{-1}
{\bf Z}$, and its only singularity on this torus is the double pole at $x=0$
which arises from the terms with $s=n_{1},...n_{M}$. The first three terms
of its Laurent decomposition can be found directly from definition,
$$W_{q}(x)=b_{0}x^{-2}+b_{1}x^{-1} +b_{2}+ O(x),$$
$$b_{0}=\exp(qn_{\beta})\prod_{\lambda\neq\beta}^{M}[\sinh \kappa(n_{\lambda}-
n_{\beta})]^{-1},$$
$$b_{1}=\kappa\{b_{0}\sum_{\gamma\neq \beta}^{M}\coth \kappa(n_{\gamma}-n_{\beta})
-\sum_{\rho\neq\beta}\exp (qn_{\rho})$$
$$\times[\sinh \kappa(n_{\beta}-n_{\rho})\prod_{\lambda\neq\rho}^{M}\sinh\kappa(n_{\lambda}-
n_{\rho})]^{-1}\},$$
$$b_{2}=\kappa^2\left\{b_{0}\left[-{1\over 3}+{{M-1}\over 2}+{1\over 2}
\sum_{\gamma\neq\delta\neq \beta}\coth(n_{\gamma}-n_{\beta})\coth (n_{\delta}-
n_{\beta})\right.\right.$$
$$\left.+\sum_{\gamma\neq\beta}\sinh^{-2}(n_{\gamma}-n_{\beta})\right]
-\sum_{\rho\neq\beta}{{\exp(qn_{\rho})}\over {\sinh \kappa (n_{\beta}-n_{\rho})}}
\prod_{\lambda\neq\rho}[\sinh \kappa(n_{\lambda}-n_{\rho})]^{-1}$$
$$\left.\times\left[\coth\kappa(n_{\beta}-n_{\rho})+\sum_{\gamma\neq\rho}
^{M}\coth\kappa(n_{\gamma}-n_{\rho})\right]\right\}+ W(p,m, \{n\}).$$
The next step consists in constructing the function $U_{q}(x)$ with the same 
quasiperiodicity and singularity at $x=0$ by using the Weierstrass functions
$\wp_{1}(x),\zeta_{1}(x)$ and\ $\sigma_{1}(x)$ defined on the torus ${\bf T}_{1}$,
$$U_{q}(x)=-A{{\sigma_{1}(x+r)}\over {\sigma_{1}(x-r)}}\exp(\delta x)$$
$$\times\{\wp_{1}(x)-\wp_{1}(r)+\Delta[\zeta_{1}(x+r)-\zeta(x)-\zeta(2r)
+\zeta(r)]\},$$
 where $A,r,\delta$ and $\Delta$ are some constants and the term in braces is chosen
as double periodic and having a zero at $x=r$. Hence the only singularity of
$U_{q}(x)$ on ${\bf T}_{1}$ is double pole at $x=0$ for all values of $r$ and
$\Delta$.

 Using the quasiperiodicity of the Weierstrass sigma function one gets
$${{\sigma_{1}(x+r+1)}\over {\sigma_{1}(x-r+1)}}=\exp(2\eta_{1}r)
  {{\sigma_{1}(x+r)  }\over {\sigma_{1}(x-r) }}, \quad
  {{\sigma_{1}(x+r+i\pi\kappa^{-1})}\over {\sigma_{1}(x-r+i\pi\kappa^{-1})}}
=\exp(2\eta_{2}r) {{\sigma_{1}(x+r)}\over{\sigma_{1}(x-r)}},$$
where $\eta_{1}=2\zeta_{1}(1/2)$ and $\eta_{2}=2\zeta_{1}(i\pi/2\kappa)$.
Comparing these expressions with quasiperiodicity of $W_{q}(x)$, one finds
two equations for $r$ and $\delta$,
$$2\eta_{1}r+\delta=-q,\quad 2\eta_{2}r+i\pi\kappa^{-1}\delta
  =i\pi(M-1).$$
Their solution can be easily found with the use of the expression for $q$
and Legendre relation $i\pi\kappa^{-1}\eta_{1}-\eta_{2}=2\pi i$,
$$r=-\left({m\over2}+{{ip}\over {4\kappa}}\right), \quad \delta=\kappa
\left[M-1+{{4i}\over\pi}r\zeta_{1}\left({{i\pi}\over{2\kappa}}\right)\right].$$
The Laurent decomposition of $U_{q}(x)$ at $x=0$ is obtained with the use of standard
expansions of the Weierstrass functions,
$$U_{q}(x)=A[x^{-2}+(2\zeta_{1}(r)+\delta-\Delta)x^{-1}+{1\over 2}(2\zeta_{1}(r)
+\delta-2\Delta)(2\zeta_{1}(r)+\delta)$$
$$+\Delta(2\zeta_{1}(r)-\zeta_{1}(2r))-\wp_{1}(r)] +O(x).$$
The function $W_{q}(x)-U_{q}(x)$ is analytic on ${\bf T}_{1}$ if $A$ and $\Delta$ obey the conditions
$$A=b_{0}, \quad A(2\zeta_{1}(r)+\delta-\Delta)=b_{1}.$$
The only analytic function which is double quasiperiodic on the torus ${\bf T}_{1}$
is zero due to the Liouville theorem. Comparison of  third terms in the decompositions of $W_{q}(x)$ and $U_{q}(x)$ gives the explicit expression of $b_{2}$
in terms of $b_{0}, r, \delta$ and $\Delta$,
$$b_{2}=b_{0}[1/2(2\zeta_{1}(r)+\delta-2\Delta)(2\zeta_{r}+\delta)+\Delta
(2\zeta_{1}(r)-\zeta_{1}(2r))-\wp_{1}(r)].$$
It allows one to find the explicit expression for the sum (48) in terms of
 $p,m,\{n\}$,
$$ W(p,m,\{n\})=\kappa^2 \left\{-\exp(qn_{\beta})\prod_{\lambda\neq\beta}^{M}
[\sinh\kappa(n_{\lambda}-n_{\beta})^{-1}\right.$$
$$\times\left[{{(M-1)}\over 2}+{1\over 2}\sum_{\gamma\neq\mu\neq\beta}^{M}
\coth\kappa(n_{\gamma}-n_{\beta})\coth\kappa(n_{\mu}-n_{\beta})\right.$$
$$\left.+\sum_{\gamma\neq\beta}[\sinh\kappa(n_{\gamma}-n_{\beta})]^{-2}-
\kappa^{-1}\tilde f(r)\sum_{\gamma\neq\beta}^{M}\coth\kappa(n_{\gamma}-n_{\beta})
+\kappa^{-2}\tilde\varepsilon(r)\right]$$ 
$$+\sum_{\rho\neq\beta}^{M}{{\exp(qn_{\rho})}\over {\sinh\kappa(n_{\beta}-n_{\rho})}}
\prod_{\gamma\neq\rho}[\sinh\kappa(n_{\gamma}-n_{\rho})]^{-1}$$
$$\left. \left[\coth\kappa(n_{\beta}-n_{\rho})+\sum_{\gamma\neq\rho}
\coth \kappa(n_{\gamma}-n_{\rho})-\kappa^{-1}\tilde f(r)\right]\right\},
\eqno(49)$$
where
$$\tilde f(r)=\zeta_{1}(2r)+\delta,$$
$$\tilde \varepsilon(r)=-{{\kappa^2}\over 3}-{1\over2}\wp_{1}(2r)+{1\over2}\tilde f
(r)^2 .$$
It is worth noting that $\tilde f$ and $\tilde\varepsilon$ are some polynomials in
$m$. Indeed, it follows from the definition of $r$ and $\delta$ that
$$r=r_{p}-{m\over 2},\quad \delta=\kappa\left[M-1-{{2i}\over \pi}m\zeta\left(
{{i\pi}\over{2\kappa}}\right)\right]+\delta_{p},$$
where
$$r_{p}=-{{ip}\over{4\kappa}},\quad \delta_{p}={p\over \pi}\zeta\left({{i\pi}\over{2\kappa}}
\right).
$$
By using quasiperiodicity of $\zeta_{1}(x)$
$$\zeta_{1}(x+l)=\zeta_{1}(x) +2l\zeta(1/2),$$
one can represent the above functions as
$$\tilde f(r)=f(p)-\kappa(2m+1-M),
$$
where
$$f(p)=\zeta_{1}(2r_{p})+\delta_{p}={p\over\pi}\zeta_{1}\left({{i\pi}\over{2\kappa}}
\right)-\zeta_{1}\left({{ip}\over {2\kappa}}\right).
\eqno(50)$$
Note that this function just coincides with the function (28) used for analysis
of two-magnon scattering. The corresponding formula for $\tilde\varepsilon$ reads
$$ \tilde\varepsilon(r)=\varepsilon(p)-\kappa(2m+1-M)f(p)+{{\kappa^2}\over 2}
(2m+1-M)^2,
\eqno(51)$$
where
$$\varepsilon(p)=-{{\kappa^2}\over 3}-{1\over 2}\wp_{1}(2r_{p})+{1\over 2}
f^{2}(p).$$
Now, according to (49-51) the left-hand side (47) of the eigenequation can be
represented as follows,
$$L(\{n\})=L_{1}(\{n\})+L_{2}(\{n\})+L_{3}(\{n\}),$$
where
$$
L_{1}(\{n\})=\psi(n_{1},...n_{M})\left[\sum_{\beta}^{M}\varepsilon (p_{\beta})
-\sum_{\beta\neq\gamma}^{M}{{\kappa^2}\over {\sinh^2\kappa(n_{\beta}-n_{\gamma})}}
\right],
$$
$$
L_{2}(\{n\})=-\kappa^2 \prod_{\mu>\nu}^{M}[\sinh\kappa(n_{\mu}-n_{\nu})]^{-1}
\sum_{P\in\pi_{M}}(-1)^{P}\sum_{m\in D^M}\tilde d_{m_{1}...m_{M}}(p)$$
$$\times\sum_{\beta\neq\rho}^{M}\exp\left[\sum_{\gamma\neq\beta, \rho}^{M}
[ip_{P\gamma}+\kappa(2m_{P\gamma}-M+1)]n_{\gamma}\right]
$$
$$\times\exp\{[i(p_{P\beta}+p_{P\rho})+2\kappa(m_{P\beta}+m_{P\rho}-M+1)]n_{\rho}
\}[\sinh\kappa(n_{\beta}-n_{\rho})]^{-1}
$$
$$\times\left[\coth\kappa(n_{\beta}-n_{\rho})+\sum_{\gamma\neq\rho}^{M}
\coth\kappa(n_{\gamma}-n_{\rho})-\kappa^{-1}f(p_{P\beta})+2m_{P\beta}-M+1\right]
$$
$$\times\prod_{\gamma\neq\beta,\rho}^{M}{{\sinh\kappa(n_{\gamma}-n_{\beta})}\over
{\sinh\kappa(n_{\gamma}-n_{\rho})}},
\eqno(52)$$
$$L_{3}(\{n\})=-\kappa^2 \prod_{\mu\neq\nu}^{M}[\sinh\kappa(n_{\mu}-n_{\nu})]^{-1}
\sum_{P\in \pi_{M}}(-1)^{P}\sum_{m\in D^M}\tilde d_{m_{1}...m_{M}}(p)
$$
$$\times\exp\left\{\sum_{\gamma=1}^{M}[ip_{P\gamma}+\kappa(2m_{P\gamma}-M+1)
]n_{\gamma}\right\}$$
$$
\left\{\sum_{\beta=1}^{M}\left[{{M-1}\over 2}-\kappa^{-1}(2m_{P\beta}-M+1)
f(p_{P\beta})+{{(M-1-2m_{P\beta})^2}\over 2}\right]\right.$$
$$-\sum_{\beta\neq\gamma}[\kappa^{-1}f(p_{P\beta})+M-1-2m_{P\beta}]\coth \kappa
(n_{\gamma}-n_{\beta})$$
$$+\sum_{\beta\neq\gamma\neq\nu}\coth\kappa(n_{\gamma}-n_{\beta})
                                \coth\kappa(n_{\nu}    -n_{\beta}).
\eqno(53)$$
Now one can see that $L_{1}(\{n\})$ exactly coincides with the right-hand side
of the equation (34) if the $M$-magnon energy is chosen as
$$\varepsilon_{M}=J\sum_{\beta=1}^{M}[\varepsilon (p_{\beta})-\varepsilon_{0}]=
J\sum_{\beta=1}^{M}\left[-{1\over 2}\wp_{1}\left({{ip_{\beta}}\over{2\kappa}}
\right)+{1\over 2}f^2(p)-{{2i\kappa}\over \pi}\zeta_{1}\left({{i\pi}\over{2a}}
\right)\right].
$$
The problem consists in finding the conditions under which $L_{2,3}(\{n\})$
vanish. Consider at first the equation (52) and denote as $Q$ the transposition
$\beta\leftrightarrow\rho$ wich does not change all other indices from 1 to $M$.
The sum over permutations in (52) can be written in the form
$$L_{2}(\{n\})=-\kappa^2\prod_{\mu>\nu}^{M}[\sinh \kappa(n_{\mu}-n_{\nu})]^{-1}
\sum_{m\in D^{M}}\tilde d_{m_{1}...m_{M}}(p)$$
$$\times \sum_{P\in\pi_{M}}(-1)^{P}\sum_{\beta\neq\rho}^{M}[F(P)-F(PQ)],$$
where 
$$ F(P)=\exp\left[\sum_{\gamma\neq \beta,\rho}^{M}(ip_{P\gamma}+\kappa(2m_{P\gamma}
-M+1))n_{\gamma}\right]$$
$$\times \exp\{[i(p_{P\beta}+p_{P\rho}+2\kappa(m_{P\beta}+m_{P\rho}-M+1)]n_{\rho}\}
$$
$$\times\sinh^{-1}\kappa(n_{\beta}-n_{\rho})\prod_{\gamma\neq\beta,\rho}^{M}
{{\sinh\kappa(n_{\gamma}-n_{\beta})}\over{\sinh\kappa(n_{\gamma}-n_{\rho})}}$$
$$\times{1\over 2}[2m_{P\beta}-\kappa^{-1}f(p_{\beta})+\coth\kappa(n_{\beta}-
n_{\rho})+\sum_{\gamma\neq\rho}\coth \kappa(n_{\gamma}-n_{\rho})-M+1].$$
Note that the only difference of $F(PQ)$ and $F(P)$ is in first two terms in
last brackets. This allows one to rewrite the last formula as
$$L_{2}(\{n\})=-\kappa^2(\prod_{\mu>\nu}[\sinh\kappa(n_{\mu}-n_{\nu})]^{-1}
\sum_{P\in\pi_{M}}(-1)^{P}$$
$$\times\sum_{\beta\neq\rho}\exp\left[\sum_{\gamma\neq\beta, \rho}[ip_{P\gamma}+\kappa
(2m_{P\gamma}-M+1)]n_{\gamma}\right]$$
$$\times \sinh^{-1}\kappa(n_{\beta}-n_{\rho})\prod_{\gamma\neq\beta,\rho}
{{\sinh\kappa(n_{\gamma}-n_{\beta})}\over{\sinh\kappa(n_{\gamma}-n_{\rho})}}$$
$$\times\sum_{\{m_{\gamma}\}\in D^{M}, \gamma\neq P \beta, P\rho}\sum_{s=0}^{2(M-1)}
\exp\{[i(p_{P\beta}+p_{P\rho})+2\kappa(s-M+1)]n_{\rho}\}$$
$$\times [M-\vert s-M+1\vert]^{-1}\sum_{m_{P\beta}+m_{P\rho}=s}\sum_{n\in{\bf Z}}
\tilde d_{m_{1}...m_{P\beta}+n...m_{P\rho}-n...m_{M}}$$
$$\times\left[m_{P\beta}-m_{P\rho}-{1\over {2\kappa}}(f(p_{P\beta})-f(p_{P\rho}))
+2n\right].$$ 
The comparison of the last sum with (42) shows that it vanishes if
$$\tilde d_{m_{1}...m_{M}}(p)=d_{m_{1}...m_{M}}(if(p)),
\eqno(54)$$
where $d_{\{m\}}(if(p))$ is an arbitrary solution to the system (20) with $p_{\mu}$
replaced by $if(p_{\mu})$, $1\leq \mu\leq M$.

The only problem is now transformation of $L_{3}(\{n\})$. Taking into account the
formula
$$\sum_{\beta\neq\gamma\neq \nu}^{M}\coth\kappa(n_{\gamma}-n_{\beta})
\coth\kappa(n_{\nu}-n_{\beta})={1\over 3}M(M-1)(M-2)$$
and symmetrizing over $\beta,\gamma$ in (53), one finds
$$L_{3}(\{n\})=-\kappa^2\prod_{\mu>\nu}[\sinh\kappa(n_{\mu}-n_{\nu})$$
$$\times\sum_{P\in\pi_{M}}(-1)^{P}\exp\left[\sum_{\gamma=1}^{M}[ip_{\gamma}-\kappa(M-1)]
n_{P^{-1}\gamma}\right]R(P,\{n\}),$$
where
$$R(P,\{n\})=\sum_{m\in D^{M}}\tilde d_{m_{1}...m_{M}}(p)\exp\left(2\kappa
\sum_{\nu=1}^{M}n_{P^{-1}\nu}m_{\nu}\right)$$
$$\times\left\{\sum_{\beta=1}^{M}\left[{1\over2}(M-1-2m_{\beta})^2 +{{M^2-1}\over 6}
-\kappa^{-1}f(p_{\beta})(2m_{\beta}-M+1)\right]\right.$$
$$-\left.\sum_{\beta\neq\gamma}[m_{\beta}-m_{\gamma}-(2\kappa)^{-1}(f(p_{\beta}-
f(p_{\gamma}))]\coth\kappa(n_{P^{-1}\beta}-n_{P^{-1}\gamma})\right\}.$$
Upon introducing the notation $\exp(2\kappa n_{P^{-1}\gamma})=y_{\gamma}$ at fixed 
$P$, one finds
$$R(P,\{n\})=\sum_{m\in D^{M}}\tilde d_{m_{1}...m_{M}}(p)\left(\prod _{\gamma=1}^{M}
y_{\gamma}^{m_{\gamma}}\right)$$
$$\times\left\{\sum_{\beta=1}^{M}\left[2m_{\beta}^2-2m_{\beta}\kappa^{-1}f(p_{\beta})-
\left(2m_{\beta}-\kappa^{-1}f(p_{\beta})-{{2M-1}\over 3}\right)(M-1)\right]\right.$$
$$\left.-\sum_{\beta\neq\gamma}^{M}{{y_{\beta}+y_{\gamma}}\over{y_{\beta}-y_{\gamma}}}
[m_{\beta}-m_{\gamma}-(2\kappa)^{-1}(f(p_{\beta})-f(p_{\gamma}))]\right\}.$$
 Now it is quite easy to see that replacing $\tilde  d_{m_{1}...m_{M}}(p)\to
d_{m_{1}..m_{M}}(p)$, $if(p_{\mu})\to p_{\mu}$ in the right-hand side just gives
the left-hand side of equation (43) and must vanish for all $y\in {\bf R}^M$
if the set $d_{\{m\}}$ solves the equation (42), i.e. the function $\chi_{p}^{(M)}$
satisfies the Calogero-Moser eigenequation. Hence both $L_{2,3}(\{n\})$ vanish under
the conditions (54) and the Ansatz (46) satisfies the eigenvalue problem (34).

These lengthy calculations lead to the simple receipt: to get a solution to (34),
one needs to change the $p$ dependence of the perodic part of the solution to
hyperbolic Calogero-Moser quantum problem as $\{p\to if(p)\}$. The asymptotic 
behaviour of the $M-$magnon wave function $\psi(n_{1},...n_{M})$ (46)as $\kappa\to
\infty$ or $\vert n_{\mu}-n_{\nu}\vert\to \infty$ can be found with the use of
Proposition 3.4. In the former case one obtains the usual Bethe Ansatz [4,5] as a
conseguence of (45) and the relation
$$\lim_{\kappa\to\infty}\kappa^{-1}[f(p_{1})-f(p_{2})]=i\left(\cot {{p_{1}}\over 2}-
\cot {{p_{2}}\over 2}\right).$$
The generalized Bethe Ansatz appears at finite $\kappa$ when the distances between
the positions of down spins tend to infinity as $n_{P(\lambda+1)}-n_{P\lambda}\to
+\infty$, $1\leq \lambda\leq M-1$,
$$\psi(n_{1},...n_{M})=\psi_{0}\sum_{Q\in \pi_{M}}(-1)^{QP}\exp\left(i\sum_{\lambda
=1}^{M}p_{Q\lambda}n_{\lambda}\right)$$
$$\times\prod_{\mu<\nu}^{M}\left\{1-{1\over {2\kappa}}[f(p_{QP\mu})-f(p_{QP\nu})
]\right\},
\eqno(55)$$
where $f(p)$ is given by the formula (50). The asymptotic form (55) will be used
in the next section within the asymptotic Bethe Ansatz scheme of calculations
of the properties of the antiferromagnetic ground state of the model.

According to (55), the multimagnon scattering matrix is factorized as it should
be for integrable models. There is a possibility for existence of multimagnon
bound complexes for which some terms in asymptotic expansion (52) vanish. 
Such a situation does not take place for usual quantum Calogero-Moser systems
with hyperbolic interaction  where the two-body potential is repulsive.\\

\noindent {\bf 4. Periodic boundary conditions and Bethe-Ansatz equations}\\

\noindent Imposing periodic boundary conditions (with period $N$) for the spin
chains with inverse square hyperbolic interaction leads to the elliptic form
of exchange (12). These conditions allows one to treat correctly also the
important case of antiferromagnetic case which corresponds to the positive 
sign of coupling constant $J$ in (12). 

The spectrum of one-magnon excitations over the ferromagnetic ground state is
now discrete and can be calculated via Fourier transform of the elliptic exchange
[17]. Throughout this section, the notation $\omega=i\pi/\kappa$ will be used
for the second period of the Weierstrass functions. As in the previous Section,
I will consider at first the case $M$=2 which allows more detailed description.

{\bf 4.1. Two-magnon scattering}. As in the case of infinite lattice, the problem
consists in finding two-magnon wave function defined by
$$\vert\psi>=\sum_{n_{1}\neq n_{2}}\psi(n_{1}, n_{2})s^{-}_{n_{1}}s_{n_{2}}^{-}
\vert 0>,$$
where $\vert\psi>$ is an eigenvector of the Hamiltonian and $\vert 0>$ is the 
"vacuum" vector with all spins up. The corresponding two-particle problem is 
now the Lame equation, and  well-known Hermite form of its solution allows one to
guess the Ansatz for the wave function in the form
$$\psi(n_{1},n_{2})={{\exp[i(p_{1}n_{1}+p_{2}n_{2})]\sigma_{N}(n_{1}-n_{2}+\gamma)
+\exp[i(p_{1}n_{2}+p_{2}n_{1})]\sigma_{N}(n_{1}+n_{2}-\gamma)}\over
{\sigma_{N}(n_{1}-n_{2})}}.$$
Since $\psi$ should be periodic in each argument, the parameters $p_{1,2}$ are
expressed through the phase $\gamma$,
$$p_{1}N-i\eta_{1}\gamma=2\pi l_{1}, \quad p_{2}N+i\eta_{1}\gamma=2\pi l_{2},
\eqno(56)$$
where $\eta_{1}=2\zeta_{N}(N/2), \eta_{2}=2\zeta_{N}(\omega/2)$ and $l_{1}, l_{2}\in {\bf Z}$.

The solution to the eigenequation is now based on the formula
$$\sum_{k=0}^{N-1}\wp_{N}(k+z){{\sigma_{N}(k-l+\gamma+z)}\over
{\sigma_{N}(k-l+z)}}\exp(i\alpha k)=-{{\sigma_{N}(l-\gamma)\sigma_{1}(z+r_
{\alpha\gamma})}\over{\sigma_{N}(l)\sigma_{1}(z-r_{\alpha\gamma})}}$$
$$\times \exp\left\{{z\over {2\pi i}}\left[\zeta_{N}(N/2)\zeta_{1}(\omega/2)
\gamma +i\zeta_{N}(\omega/2)\alpha\right]\right\}$$
$$\times \{\wp_{1}(z)-\wp_{1}(r_{\alpha\gamma})+2(\zeta_{1}(z+r_{\alpha\gamma})
-\zeta_{1}(z) +\zeta_{1}(r_{\alpha\gamma})-\zeta_{1}(2r_{\alpha\gamma}))$$
$$\times [\zeta_{1}(r_{\alpha\gamma})+{{\zeta_{N}(l-\gamma)-\zeta_{N}(l)}\over 2}
-{{\exp(i\alpha l)\sigma_{N}(\gamma)\sigma_{N}(l)}\over{2\sigma_{N}(l-\gamma)}}
\wp_{1}(l)$$
$$+{1\over 4\pi i}(\zeta_{N}(N/2)\zeta_{1}(\omega/2)\gamma+i\zeta_{N}(\omega/2)\alpha
)]\},$$
where $l\in {\bf Z}$ and $\alpha$ and $\gamma$ are connected by
$$\exp[i\alpha N+ 2\gamma\zeta_{N}(N/2)]=1, \quad r_{\alpha\gamma}=-(4\pi)^{-1}
[\alpha\omega-i\gamma\zeta_{N}(\omega/2)].$$
The two-magnon energy is given by
$$\varepsilon_{2}(p_{1},p_{2},\gamma)=J\{1/4[f(p_{1}, \gamma)+f(p_{2},-\gamma)]^{2}
+\varepsilon_{0}(p_{1}, \gamma)+\varepsilon_{0}(p_{2},-\gamma)+\wp(\gamma)\},
$$
where
$$\varepsilon_{0}(p,\gamma)={2\over \omega}[\zeta_{1}(\omega/2)-N\zeta_{N}(\omega/2)]
-{1\over 2}\wp_{1}\left({{i\eta_{2}\gamma-p\omega}\over {2\pi}}\right),$$,
$$f(p,\gamma)=\zeta_{1}\left({{i\eta_{2}\gamma-p\omega}\over{2\pi}}\right)
+(i\pi)^{-1}\left[\eta_{2}\zeta_{1}(1/2)+ip\zeta_{1}(\omega/2)\right]
\eqno(57)$$
and $p_{1,2}$ and $\gamma$ are constrained by
$$f(p_{1},\gamma)-f(p_{2}, -\gamma)-2\zeta_{N}(\gamma)=0.
\eqno(57)$$
With the use of (56),(57) and direct computation it is possible to show that
$$S^{+}\sum_{n{1}\neq n_{2}}^{N}\psi(n_{1}, n_{2})s_{n_{1}}^{-}s_{n_{2}}^{-}\vert 0>=0,$$
i.e. these states have the total spin $S=S_{z}=N/2-2$.

It is natural to ask of how many solutions do the equations (56), (57) have.
The completeness of the set of these solutions means that their number should
be equal $N(N-3)/2$ since in two-magnon sector there are $N$ solutions with
$\psi(n_{1}, n_{2})=\psi_{1}(n_{1})+\psi_{1}(n_{2})$. Is it possible to evaluate
the number of solutions to (56),(57) analytically? The answer is positive [22].
The sketch of the proof is as follows. The constraint (57) can be rewritten as
$$F_{l_{1}, l_{2}}(\gamma)=\zeta_{1}\left({{\gamma-l_{1}\omega}\over N}\right)
                          +\zeta_{1}\left({{\gamma+l_{2}\omega}\over N}\right)
+2{{l_{1}-l_{2}}\over N}\zeta_{1}(\omega/2)$$
$$+{{4\gamma}\over\omega}[\zeta_{N}(\omega/2)-N^{-1}\zeta_{1}(\omega/2)]-2\zeta_{N}
(\gamma)=0.$$
At fixed $l_{1,2}$, it is a transcendental equation for $\gamma$.

Let now $\Lambda$ be the manifold which consists of various sets $\{l_{1,2}\in {\bf Z},
\gamma\in{\bf C}\}$ and call two sets $\{l_{1},l_{2},\gamma\}$, $\{l'_{1},l'_{2},
\gamma'\}\in\Lambda$ equivalent if the corresponding wave functions coincide up
to normalization factor. With the use of (56) and quasiperiodicity of sigma functions,
one finds that the manifold $\Lambda$ is equivalent to its submanifold $\Lambda_{0}$
defined by the relations
$$0\leq l_{1}\leq N-1, \quad l_{2}=0, \quad \gamma\in {\bf T}_{N,N\omega}.$$
Let $\{\lambda\}$ be a variety of nonequivalent sets within $\Lambda_{0}$. 
The question now is: how many nonequivalent sets obeying $F_{l_{1},0}(\gamma)=0$ are in $\Lambda_{0}$? To answer it, let us note that the function $F_{l_{1},0}(\gamma)$ is
double periodic with periods $N$ and $N\omega$ and there is the relation between
$\zeta $ functions of periods $(N,\omega)$ and $(N,N\omega)$,
$$\zeta_{N}(x)=\zeta(x)+\sum_{j=1}^{N-1}[\zeta(x+j\omega)-\zeta(j\omega)]
+{{2x}\over \omega}[\zeta_{N}(\omega/2)-\zeta(N\omega/2)],$$
where $\zeta(x)$ is the zeta function defined on the torus ${\bf T}_{N,N\omega}$.
With the use of scaling relation $\zeta_{1}(N^{-1}(x)=N\zeta(x)$ one can rewrite
 the constraint $F_{l_{1},0}(\gamma)=0$ in the form
$$-2\sum_{j=0}^{N-1}\zeta(\gamma-j\omega)+N[\zeta(\gamma-l_{1}\omega)+\zeta(\gamma)]
+2\zeta\left({{N\omega}\over 2}\right)(l_{1}-N+1)=0.
\eqno(58)$$
It is easy to see that at $N>2$ there are $N$ simple poles of the left-hand side of
equation (58) located at $\gamma=j\omega$ and this function is elliptic. Then there
should be just $N$ roots of equation (58) within ${\bf T}_{N,N\omega}$ and, at first
sight, $\{\lambda\}$ consists of $N^2$ elements.  However, some sets with different
roots of (58) may be equivalent. In fact, one can see that if $\gamma_{0}\in {\bf T}
_{N,N\omega}$ is a root, then
$$\gamma'_{0}=-\gamma_{0}+l_{1}\omega+N{\rm sign}(\Re e \gamma_{0})+{{N\omega}\over 2}
[1-{\rm sign} (l_{1}\vert\omega\vert-\Im m \gamma_{0})]$$                                    is also a root of (58). Moreover, all the solutions to the equation $\gamma_{0}'=
\gamma_{0}$ are the roots of (58). The sets of these solutions are different 
for $N$, $l_{1}$ even or odd. There are four cases. If both $N$ and $\omega$ are even, there are only two these roots, $(N+l_{1}\omega)/2$ and $(N+l_{1}\omega+N\omega)/2$.
 If both $N$ and $l_{1}$
are odd, the additional root $l_{1}\omega/2$ is present. As $N$ is odd and $l_{1}$
even, the additional root is $(N+l_{1})\omega/2$. And in the case of even $N$ and
odd $l$, one has four such a roots since both $l_{1}\omega/2$ and $(N+l_{1})\omega/2$
obey the equation (58). 

All these explicit roots are combinations of half-integer periods of the torus
${\bf T}_{N,N\omega}$. In this case the wave function can be simplified and it
turns out that $\psi(n_{1}, n_{2})$ vanishes identically for all explicit roots
listed above.

The number of all nontrivial and nonequivalent sets $\{l_{1},0,\gamma\}$ in the
variety $\{\lambda\}$ can be now easily counted. At even $N$, there are $N/2$
even $\{l_{1}\}$ with $(N/2)-1$ nonequivalent roots and $N/2$ odd $\{l_{1}\}$
with $(N/2)-2$ ones. At odd $N$, there are $(N+1)/2$ even $\{l_{1}\}$ and
$(N-1)/2$ odd $\{l_{1}\}$ with $(N-3)/2$ nonequivalent roots in both cases.
Hence the total number of elements in the variety $\{\lambda\}$ equals $N(N-3)/2$ 
as it should be, and the nonequivalent solutions to (58) provide complete
description of nontrivial two-magnon states. It would be of interest to investigate,
in the limit of large $N$, the distribution of nontrivial roots within the torus
${\bf T}_{N,N\gamma}$.

{\bf 4.2. Multimagnon states}. As in preceding Section, one has to investigate first
the solutions to usual quantum Calogero-Moser problem with coupling constant $l=1$
in (3) and elliptic two-body potential. This problem at $M>2$ was attacked first
in [24] where the general statements on the structure of many-particle wave function
have been proved and explicit result for $M=3$ has been obtained. The problem of
arbitrary $l$ and $M=3$ has been considered in [25] and soon the analytic
expression for arbitrary $M$ has been also found [39] in the process of solving
the elliptic Knizhnik-Zamolodchikov-Bernard equation. Unfortunately, its form
turned out to be so complicated that no explicit calculation were possible for
multimagnon wave functions. At $M=3$, the 3-magnon wave function has been found
explicitly in [31] but the calculations were very lengthy and it has been not seen
how to generalize the method for $M>3$. The way to the solution of the $M$-magnon
problem which does not refer to explicit form of the solution to $M$-particle
problem has been found later [30,40]. Before describing it, it will be of use
to formulate basic facts about the wave functions of the continuous $M$-particle problem for elliptic two-body interaction [24].

Since $\wp(x)$ is double periodic, it is easy to see that the corresponding $M$-
particle Hamiltonian (3) commutes with $2M$ shift operators $Q_{\alpha j}=
\exp(\omega_{\alpha}\partial/\partial x_{j})$, where $\omega_{1,2}$ are two
periods of $\wp(x)$. Let $\chi^{(p)}(x_{1},...x_{M})$ be their common eigenvector,
$$\chi^{(p)}(x_{1}+\sum_{\alpha=1}^2 l_{1}^{(\alpha)}\omega_{\alpha},...
x_{M}+\sum_{\alpha=1}^{2}l^{(\alpha)}_{M}\omega_{\alpha})= \exp(i\sum_{j=1}^{M}
\sum_{\alpha=1}^{2}p_{\alpha}^{(j)}l_{j}^{(\alpha)})\chi^{(p)}(x_{1},...x_{M}),
$$
where $l_{j}^{(\alpha)}\in {\bf Z}$. Hence $\chi^{(p)}(x)$ can be treated
on the $M$-dimensional torus ${\bf T}_{M}={\bf C}/{\bf Z}\omega_{1}+{\bf Z}
\omega_{2}$ with quasiperiodic boundary conditions. The structure of singularities
of the Hamiltonian (3) in this torus shows that $\chi^{(p)}$ is analytic except of
all hypersurfaces $L_{jk}$ defined by the equalities $x_{j}=x_{k}$, $1\leq j<k\leq M$.
On each $L_{jk}$, $\chi^{(q)}$ has a simple pole. Let $\Psi_{M}$ be a class of functions with these properties.\\
{\bf Proposition 4.1.} The class $\Psi_{M}$ is a functional manifold of dimension
$2M-1+M^{M-2}$. The parameters $\{p_{\alpha}^{(j)}\}$ are not independent but restricted
by the linear relation $\sum_{j=1}^{M}(p_{1}^{(j)}\omega_{2}-p_{2}^{(j)}\omega_{1})=
{\bf Z}\omega_{1}+{\bf Z}\omega_{2}$. The manifold $\Psi_{M}$ can be described as an
union of the $(2M-1)$-parametric family of linear spaces with dimensions $M^{M-2}$
with the basic vectors parametrized by $\{p_{\alpha}^{(j)}\}$.\\
{\bf Proposition 4.2}. The co-ordinate system on $\Psi_{M}$ can be chosen by such a
way that all its elements are expressed through the Riemann theta functions of genus 1
or usual Weierstrass sigma functions.\\
The sketch of the proofs can be found in [24]. The explicit expressions for $\chi^{(p)} $ can be also found in [24] for $M=3$ and in [39] for arbitrary $M$. The amazing fact
is that the treatment of $M$-magnon problem can be done {\it without} use of these
explicit expressions.

Let us choose the exchange in the form
$$h(j)=J\left({\omega\over\pi}\sin{\pi\over\omega}\right)^2
\left[\wp_{N}(j)+{2\over\omega}\zeta_{N}\left({\omega\over2}\right)\right]
$$
 so as to reproduce correctly the inverse square hyperbolic form of the Section 3
in the thermodynamic limit $N\to\infty$. The second period of the Weierstrass
function $\wp_{N}$ is $\omega=i\pi/\kappa$. The eigenproblem is decomposed into
the problems with $M$ down spins due to rotation invariance and the eigenvectors
$\vert\psi^{(M)}>$ are given by
$$\vert\psi^{(M)}>=\sum_{n_{1}..n_{M}}^{N}\psi_{M}(n_{1}..n_{M})
\prod_{\beta=1}^{M}
s_{n_{\beta}}^{-}\vert0>,
$$
where $\vert0>=\vert\uparrow\uparrow...\uparrow>$ is the ferromagnetic ground state with all spins up and the summation is taken over all combinations of integers $\{n\}\leq N$ such that
$\prod_{\mu<\nu}^{M}(n_{\mu}-n_{\nu})\neq0$. The function $\psi_{M}$ is completely
symmetric in its arguments and obeys lattice Schr\"odinger equation
$$\sum_{s\neq n_{1},..n_{M}}^{N}\sum_{\beta=1}^{M}\wp_{N}(n_{\beta}-s)
\psi_{M}(n_{1},..n_{\beta-1},s,n_{\beta+1},..n_{M})$$
$$+
\left[\sum_{\beta\neq\gamma}^{M}\wp_{N}(n_{\beta}-n_{\gamma})-{\cal E}_{M}
\right]\psi_{M}(n_{1},..n_{M})=0
\eqno(59)$$
and the eigenvalues of the Hamiltonian are given by
$$\varepsilon_{M}=J\left({\omega\over\pi}\sin{\pi\over\omega}\right)^{2}
\left\{{\cal E}_{M}+{2\over\omega}\left[{{2M(2M-1)-N}\over4}\zeta_{N}\left(
{\omega\over2}\right)-M\zeta_{1}\left({\omega\over2}\right)\right]
\right\}.
$$
Let $\chi_{M}^{(p)}$ be the special solution to the continuum quantum many-particle
problem 
$$\left[-{1\over2}\sum_{\beta=1}^{M}{{\partial^2}\over{\partial x_{\beta}^2}}
+\sum_{\beta\neq\lambda}^{M}\wp_{N}(x_{\beta}-x_{\lambda})-{\sf E}_{M}(p)\right]
\chi_{M}^{(p)}(x_{1},..x_{M})=0,
$$
which is specified up to some normalization factor by particle pseudomomenta 
$(p_{1},...p_{M})$ and obeys the quasiperiodicity conditions
$$\chi_{M}^{(p)}(x_{1},..x_{\beta}+N,..x_{M})=\exp(ip_{\beta}N)
\chi_{M}^{(p)}(x_{1},..x_{M}),\quad 1\leq\beta\leq M,
$$
$$\chi_{M}^{(p)}(x_{1},..x_{\beta}+\omega,..x_{M})=\exp
(2\pi iq_{\beta}(p)+ip_{\beta}\omega)
\chi_{M}^{(p)}(x_{1},..x_{M}), \quad 0\leq{\Re}e(q_{\beta})<1.
\eqno(60)$$
As will be seen later, the set $\{q_{\beta}(p)\}$ is 
completely determined by $\{p\}$.

The connection of $\chi_{M}^{(p)}$ with multimagnon wave function is given by the 
Ansatz$$
\psi_{M}(n_{1},..n_{M})=\sum_{P\in\pi_{M}}\varphi_{M}^{(p)}(n_{P1},..n_{PM}),
$$
$$\varphi_{M}^{(p)}(n_{1},..n_{M})=\exp\left(-i\sum_{\nu=1}^{M}\tilde p_{\nu}
n_{\nu}
\right)
\chi_{M}^{(p)}(n_{1},..n_{M}),
\eqno(61)$$
where
$$\tilde p_{\nu}=p_{\nu}-2\pi N^{-1}l_{\nu}, \qquad l_{\nu}\in {\bf Z}.
$$
The last condition is just the condition of periodicity of $\psi_{M}$. The problem
now consists in calculation of the left-hand side of the lattice Schr\"odinger 
equation (59), but before doing this let us mention that $\chi_{M}{(p)}$ has
the singularities in the form of simple poles and can be presented in the form
$$\chi_{M}^{(p)}={{F^{(p)}(x_{1},..x_{M})}\over{G(x_{1},..x_{M})}},\quad
G(x_{1},..x_{M})=\prod_{\alpha<\beta}^{M}
\sigma_{N}(x_{\alpha}-x_{\beta}),
\eqno(62)$$
where $\sigma_{N}(x)$ is the Weierstrass sigma function on the torus $T_{N}$.
By definition,
the only simple zero of $\sigma_{N}(x)$ on $T_{N}$ is located at $x=0$. Thus
$[G(x_{1},..x_{M})]^{-1}$ absorbs all the singularities of $\chi_{M}^{(p)}$
on the hypersurfaces $x_{\alpha}=x_{\beta}$. The numerator $F^{(p)}$ in (62)
should be analytic on $(T_{N})^M$. It obeys the equation
$$
\sum_{\alpha=1}^{M}{{\partial^2 F^{(p)}}\over{\partial x_{\alpha}^2}}
+\left[2E_{M}(p)-{M\over2}\sum_{\alpha\neq\beta}^{M}
(\wp_{N}(x_{\alpha}-x_{\beta})-\zeta_{N}^{2}(x_{\alpha}-x_{\beta}))\right]F^{(p)}
$$
$$=\sum_{\alpha\neq\beta}\zeta_{N}(x_{\alpha}-x_{\beta})
\left({{\partial F^{(p)}}\over{\partial x_{\alpha}}}-
      {{\partial F^{(p)}}\over{\partial x_{\beta}}}\right).
$$
The left-hand side of this equation is regular as
$x_{\mu}\to x_{\nu}$. Hence $F^{(p)}$ must obey the condition

$$\left({\partial\over{\partial x_{\mu}}}-{\partial\over{\partial x_{\nu}}}\right)
F^{(p)}(x_{1},..x_{M})\vert_
{x_{\mu}=x_{\nu}}=0
\eqno(63)$$
for any pair $(\mu,\nu)$. Let us now show that the properties (60), (62,63) allow
to validate the ansatz (61) for $\psi_{M}$. Substitution of (61) to (59) yields
$$\sum_{P\in\pi_{M}}\left\{\sum_{\beta=1}^{M}{\cal S}_{\beta}(n_{P1},..n_{PM})
+\left[\sum_{\beta\neq\gamma}^{M}\wp_{N}(n_{P\beta}-n_{P\gamma})-{\cal E}_{M}
\right]\varphi^{(p)}_{M}(n_{P1},..n_{PM})\right\}=0,
$$
where
$${\cal S}_{\beta}(n_{P1},..n_{PM})=\sum_{s\neq n_{P1},..n_{PM}}^{N}
\wp_{N}(n_{P\beta}-s)\hat Q_{\beta}^{(s)}\varphi_{M}^{(p)}
(n_{P1},..n_{PM})
\eqno(64)
$$
and the operator $\hat Q_{\beta}^{(s)}$  replaces $\beta$th  argument
of the function of $M$ variables to $s$.

The calculation of the sum (64) is based on introducing the function of complex
variable $x$
$$W_{P}^{(\beta)}(x)=\sum_{s=1}^{N}\wp_{N}(n_{P\beta}-s-x)\hat Q_{\beta}^{(s+x)}
\varphi_{M}^{(p)}(n_{P1},..n_{PM}).
$$
As a consequence of (60) it obeys the relations
$$W_{P}^{(\beta)}(x+1)=W_{P}^{(\beta)}(x),\qquad
W_{P}^{(\beta)}(x+\omega)=\exp(2\pi i\tilde q_{\beta}(p))W_{P}^{(\beta)}
(x),
\eqno(65)$$
where
$$\tilde q_{\beta}(p)= q_{\beta}(p)+{{l_{\beta}}\over N}\omega.$$
The only singularity of $W_{P}^{(\beta)}$ on the torus $T_{1}={\bf C}/{\bf Z}+
{\bf Z}\omega$ is located at the point $x=0$. It arises from the terms in (64)
with $s=n_{P1},..n_{PM}$. Hence the Laurent decomposition of $W_{P}^{(\beta)}$ near $x=0$
has the form
$$W_{P}^{(\beta)}(x)=w_{-2}x^{-2}+w_{-1}x^{-1}+w_{0}+O(x).
\eqno(66)$$
With the use of (62), one can find the explicit expressions for $w_{-i}$ in the form
$$w_{-2}=\varphi_{M}^{(p)}(n_{P1},..n_{PM})
$$
$$w_{-1}={\partial\over{\partial n_{P\beta}}}\varphi_{M}^{(p)}(n_{P1},..n_{PM})
$$
$$
+(-1)^P G^{-1}(n_{1},..n_{M})\sum_{\lambda\neq\beta}T_{\beta\lambda}
(n_{P1},..n_{PM})
\hat Q_{\beta}^{(n_{P\lambda})}
\exp\left(-i\sum_{\nu=1}^M \tilde p_{\nu}n_{P\nu}\right)F^{(p)}(n_{P1},..n_{PM})
$$
$$w_{0}={\cal S}_{\beta}(n_{P1},..n_{PM})+{1\over2}{{\partial^2}\over
{\partial n_{P\beta}^2}}\varphi_{M}^{(p)}(n_{P1},..n_{PM})
+(-1)^{P}G^{-1}(n_{1},..n_{M})$$
$$\times\sum_{\lambda\neq\beta}T_{\beta\lambda}
(n_{P1},..n_{PM})\left[U_{\beta\lambda}(n_{P1},..n_{PM})
\hat Q_{\beta}^{(n_{P\lambda})}
+\wp_{N}(n_{P\beta}-n_{P\lambda})\partial \hat Q_{\beta}^{(n_{P\lambda})}
\right]
$$
$$\times\exp\left(-i\sum_{\nu=1}^M \tilde p_{\nu}n_{P\nu}\right)
F^{(p)}(n_{P1},..n_{PM}),$$
where
$$T_{\beta\lambda}(n_{P1},..n_{PM})=\sigma_{N}(n_{P\lambda}-n_{P\beta})
\prod_{\rho\neq\beta,\lambda}^{M}
{{\sigma_{N}(n_{P\rho}-n_{P\beta})}\over
{\sigma_{N}(n_{P\rho}-n_{P\lambda})}},$$
$$
U_{\beta\lambda}(n_{P1},..n_{PM})=\wp_{N}'(n_{P\lambda}-n_{P\beta})-
\wp_{N}(n_{P\beta}-n_{P\lambda})
\sum_{\rho\neq\beta,\lambda}\zeta_{N}(n_{P\rho}-n_{P\lambda}),$$
$(-1)^{P}$ is the parity of the permutation $P$ and the action of the operator
 $\partial\hat Q_{\beta}^{(n_{P\lambda})}$
on the function $Y$ of $M$ variables is defined as
$$\partial Q_{\beta}^{(n_{P\lambda})}Y(z_{1},..z_{M})=
{\partial\over{\partial z_{\beta}}}Y(z_{1},..z_{M})\vert_{z_{\beta}=
n_{P\lambda}}.
$$
Note now that the expression
for the function $W_{P}^{(\beta)}(x)$ obeying the relations
(65) and (66) can be written analytically without any further freedom,
$$W_{P}^{(\beta)}(x)=\exp(a_{\beta}x)
{{\sigma_{1}(r_{\beta}+x)}\over
{\sigma_{1}(r_{\beta}-x)}}\{w_{-2}(\wp_{1}(x)-\wp_{1}(r_{\beta}))+
(w_{-2}(a_{\beta}+2\zeta_{1}(r_{\beta}))-w_{-1})
$$
$$
\times[\zeta_{1}(x-r_{\beta})-\zeta_{1}(x)+\zeta_{1}(r_{\beta})-
\zeta_{1}(2r_{\beta})]\}.
$$
The parameters $a_{\beta},r_{\beta}$
are chosen as to satisfy the conditions (65),
$$a_{\beta}=2\tilde q_{\beta}(p)\zeta_{1}(1/2),
\qquad r_{\beta}=-{1\over 2}\tilde q_{\beta}(p).$$
By expanding the above form of $W^{(\beta)}_{P}$ in powers of $x$ one can find $w_{0}$ in terms of $w_{-2},w_{-1},
q_{\beta}$ and obtain the explicit expression for
${\cal S}_{\beta}(n_{P1},..n_{PM})$ . After long but
straightforward calculations the equation (59) can be recast in the form
$$
\sum_{P\in\pi_{M}}\left[-{1\over2}\sum_{\beta=1}^{M}
\left({\partial\over{\partial n_{P\beta}}}-f_{\beta}(p)\right)^2+
\sum_{\beta\neq\gamma}^{M}\wp_{N}(n_{P\beta}-n_{P\gamma})-{\cal E}_{M}+
\sum_{\beta=1}^{M}\varepsilon_{\beta}(p)\right]\varphi^{(p)}(n_{P1},..
n_{PM})$$
$$={1\over2}G^{-1}(n_{1},..n_{M})\sum_{P\in\pi_{M}}(-1)^{P}
\sum_{\beta\neq\lambda}\left[
Z_{\beta\lambda}(n_{P1},..n_{PM})+Z_{\lambda\beta}(n_{P1},..n_{PM})
\right],
\eqno(67)
$$
where
$$f_{\beta}(p)=2\tilde q_{\beta}(p)\zeta_{1}(1/2)
-\zeta_{1}(\tilde q_{\beta}(p)),
$$
$$\varepsilon_{\beta}(p)={1\over2}\wp_{1}(\tilde q_{\beta}(p))
$$
and $Z_{\beta\lambda}(n_{P1},..n_{PM})$ is defined by the relation
$$Z_{\beta\lambda}(n_{P1},..n_{PM})=T_{\beta\lambda}(n_{P1},..n_{PM})
\left[U_{\beta\lambda}(n_{P1},..n_{PM})
\hat Q_{\beta}^{(n_{P\lambda})}+\wp_{N}(n_{P\lambda}-n_{P\beta})\right.
$$
$$\left.\times
(\partial\hat Q_{\beta}^{(n_{P\lambda})}-f_{\beta}(p)
        \hat Q_{\beta}^{(n_{P\lambda})})\right]
\exp\left(-i\sum_{\nu=1}^{M}\tilde p_{\nu}n_{P\nu}\right)
F^{(p)}(n_{P1},..n_{PM}).
$$
One observes with the use of the definition (61) of $\varphi^{(p)}$,
 that each term of the left-hand side of (67) has the same structure
as the left-hand side of the many-particle Schr\"odinger equation and
vanishes if ${\cal E}_{M}$ and $f_{\beta}(p)$ are chosen as
$$f_{\beta}(p)=-i\tilde p_{\beta},\qquad \beta=1,..M,
\eqno(68a)$$
$${\cal E}_{M}={\sf E}_{M}(p)+\sum_{\beta=1}^{M}\varepsilon_{\beta}(p).
\eqno(68b)$$
It remains to prove that that the right-hand side of (67) also vanishes.
It can be done by using the observation that the sum over permutations in it
can be simply recast in the form
$$\sum_{P\in\pi_{M}}(-1)^{P}\sum_{\beta\neq\lambda}[Z_{
\beta\lambda}(n_{P1},..n_{PM})-Z_{\lambda\beta}(n_{PR1},..n_{PRM})],$$
where $R$ is the transposition $(\beta\leftrightarrow\lambda)$
which leaves other numbers from 1 to $M$ unchanged.
Taking into account the fefinition of $Z$,
one finds
$$Z_{\beta\lambda}(n_{P1},..n_{PM})-Z_{\lambda\beta}(n_{PR1},..n_{PRM})=
T_{\beta\lambda}(n_{P1},..n_{PM})\wp_{N}(n_{P\lambda}-n_{P\beta})$$
$$\times\exp\left[-i\left((\tilde p_{\beta}+\tilde p_{\lambda})n_{P\lambda}+
\sum_{\rho\neq\beta,\lambda}^{M}\tilde p_{\rho}n_{P\rho}
\right)\right]\left(
{\partial\over{\partial n_{P\beta}}}-
{\partial\over{\partial n_{P\lambda}}}\right)F^{(p)}
(n_{P1},..n_{PM})\vert_{n_{P\beta}=n_{P\lambda}}.
$$
Now it is clearly seen that the last factor  vanishes due to the
condition (63). 

The relations (68a-b) for the spectrum are still not complete since
the dependence of $\{q\}$ on $\{p\}$ is not known on this stage. This
 completion can be done only by further analysis of the properties of
$\chi_{M}^{(p)}$ solving $M$-particle Schr\"odinger equation.
In [39]  the explicit form of $\chi_{M}^{(p)}(x)$ has been
found in the process of solving the Knizhnik-Zamolodchikov-Bernard
 equations. In suitable notations, it reads
$$\chi_{M}^{(p)}(x)\sim\exp(i\sum_{\beta=1}^{M}p_{\beta}x_{\beta})
\sum_{s\in \pi_{m}}l(s)\prod_{j=1}^{m}\tilde\sigma_{\sum_{k=1}^j (x_{c(s(k))
}-x_{c(s(k))+1})}(t_{s(j)}-t_{s(j+1)}),
\eqno(69)$$
where $m=M(M-1)/2$, $c$ is non-decreasing function $c:\{1,..,m\}\to
\{1,..,M-1\}$ such that $\vert c^{-1}\{j\}\vert=M-j$,
  $l(s)$ is an integer which is defined for the
permutation $s$ by the relation $x_{c(s(1))+1}\partial/\partial x_{c(s(1))}
...x_{c(s(m))+1}\partial/\partial x_{c(s(m))}x_{1}^{M}=$ $l(s)(x_{1}...x_{M})
$, $\{t\}$ is a set of $m$ complex parameters obeying $m$ relations [39]
$$\sum_{l:\vert c(l)-c(j)\vert =1}\rho(t_{j}-t_{l})-2\sum_{l:l\neq j,c(l)=
c(j)}\rho(t_{j}-t_{l})+M\delta_{c_{j},1}\rho(t_{j})=i(p_{c(j)}-p_{c(j)+1}),
\eqno(70)$$
$$\rho(t)=\zeta_{N}(t)-{2\over N}\zeta_{N}(N/2)t,$$
and 
$$\tilde\sigma_{w}(t)=\exp((2/N)\zeta_{N}(N/2)wt){{\sigma_{N}(w-t)}\over
{\sigma_{N}(w)\sigma_{N}(t)}},$$
The main advantage of the explicit form of $\chi$ function is that it
allows to find the second set of relations between the Bloch factors
$\{p\}, \{q\}$. It is easy to see that $\{p\}'s$ in the definitions
(60) and (69) are the same. The problem consists in calculation of $\{q\}$.
To do this, it is not necessary to analyze each term in the sum over 
permutations in (69) since all of them must have the same Bloch factors.
It is convinient to choose the term which corresponds to the permutation 
$$s_{0}: \quad s_{0}(j)=m+1-j,\quad j=1,..m.$$
After some algebra, one finds that this permutation gives nontrivial
contribution to (69) with $l(s_{0})=M!(M-1)!...2!$. Moreover,
with the use of explicit form of the color function  one finds
$$c(s_{0}(l))=M-q\quad {\rm if}\quad q(q-1)/2+1\leq l\leq q(q+1)/2.$$
 Now the problem of calculation of the second Bloch factors reduces
 to some long and tedious, but in fact simple 
calculations of the product of factors which various $\tilde\sigma$ functions
acquire under changing arguments of $\chi$ function to the quasiperiod
 $\omega$. The final result is surprisingly simple,
$$q_{\beta}(p)=N^{-1}\left(\sum_{l:c(l)=\beta}t_{l}-\sum_{l:c(l)=\beta -1}
t_{l}\right),\qquad 1<\beta<M-1,
\eqno(71)$$
with the first and second term being omitted for $\beta=M$ and $\beta=1$.
 
The equations (71), together with (68a) and (70), form a closed set
for finding Bloch factors $\{p\},\{q\}$ at given integers
$\{l_{\beta}\}\in {\bf Z}/M{\bf Z}$ and determining the eigenvalues of
the spin Hamiltonian completely.
The corresponding eigenvalue of the continuum $M$-particle operator  is
given by [39]
$${\sf E}_{M}(p)={{2M(M-1)}\over N}\zeta_{N}\left({N\over 2}\right)
+\sum_{\beta=1}^{M}p_{\beta}^{2}/2$$ 
$$-{1\over 2}\left[\sum_{k<l}^{m}(2\delta_{c(k),c(l)}F(t_{k}-t_{l})-
\delta_{\vert c(k)-c(l)\vert,1}F(t_{k}-t_{l}))-M\sum_{c(k)=1}F(t_{k})
\right],$$
where
$$
F(t)=-\wp_{N}(t)+(\zeta_{N}(t)-2/N\zeta_{N}(N/2))^2 +4/N\zeta_{N}(N/2).
$$
It is worth noting that for
real calculation of the eigenvalues one has to solve the Bethe-type equations (68a), (70),(71) at first. It is not clear how to treat properly this huge
system of highly transcendental equations even in the limit $N\to\infty$. In this
limit, there is a procedure known as asymptotic Bethe Ansatz  (ABA) which consists in
imposing periodic boundary condition on the asymptotics  of the
wave functions for infinite lattice [41]. It will be used in the next subsection
for obtaining some results on antiferromagnetic ground state.

{\bf 4.3. ABA results for large N}. In this subsection, the ABA hypothesis (still
unproved) will be used for description of some properties of the spin chain with
the exchange
$$h(j)=
{{\sinh^2{\kappa}}\over{\sinh^2 \kappa(j-k)}},
\eqno(72)$$
which corresponds to $J=-(\sinh \kappa/\kappa)^2$ in (23) (the antiferromagnetic
regime) at large but finite $N$. Note that in the nearest-neighbor limit
$\kappa\to\infty$ one can decompose (23) with the exchange (72) as
$$ H= {1\over 2}\sum_{j}(\vec \sigma_{j}\vec \sigma_{j+1}
-1)+{1\over 2}e^{-2\kappa}\sum_{j}(\vec \sigma_{j}\vec \sigma_{j+2}-1)
+o(e^{-2\kappa}).$$
Hence one can write the ground-state energy per site as
$$e={1\over 2}<\vec \sigma_{j}\vec \sigma_{j+1}-1>
+{1\over 2}e^{-2\kappa}<\vec \sigma_{j}\vec \sigma_{j+2}-1> +o(e^{-2a}),
\eqno(73)
$$
where $<$ $>$ means average on the vacuum state of the Hamiltonian.
Fortunately, in the first order approximation  one can replace this state
to the vacuum state of nonperturbed Hamiltonian with the interaction
of nearest-neighbor spins, $H_{0}={1\over 2}\sum_{j}
(\vec \sigma_{j}\vec \sigma_{j+1}-1)$. It gives an opportunity to find the 
second-neighbor correlator $\vec \sigma_{j}\vec\sigma_{j+2}$ by calculating
(73) explicitly.

The scheme of ABA is based on asymptotic expression of the wave function with
$M$ down spoins for infinite chain in the region
$n_{1}\ll n_{2}...\ll n_{M}$, which has been described in Section 3,
$$\psi(n_{1},..n_{M})\propto\sum_{P\in \pi_{M}}
\exp\left(i\sum_{\alpha=1}^M k_{P\alpha}n_{\alpha}\right)
\exp\left({i\over 2}\sum_{\alpha<\beta}^M \chi(p_{P\alpha},p_{P\beta})\right
), $$
where the first sum is taken over all permutations from the group $\pi_{M}$,
$\{p_{\alpha}\}$ is the set of pseudomomenta and $\chi(p_{\alpha},p_{\beta})$
is the two-magnon phase shift defined by the relations
$$\cot{{\chi(p_{\alpha},p_{\beta})}\over 2}=\varphi(p_{\alpha})-
\varphi(p_{\beta}),$$
 $$\varphi(p)={p\over{2\pi i \kappa}}\zeta_{1}\left({{i\pi}\over{2\kappa}}\right)-
{1\over {2i\kappa}}\zeta_{1}\left({{ip}\over {2\kappa}}\right).$$
To consider the chains of finite length $N$ in the thermodynamic limit
$N\to\infty$, we adopt the main hypothesis of ABA, i.e. imposing periodic
boundary conditions on the asymptotic form of the wave function.
Taking $\psi(n_{2},..n_{M},n_{1}+N)=\psi(n_{1},..n_{M})$ and calculating 
the both  sides
with the use of the above formula for asymptotics
results in the ABA equations
$$\exp(ip_{\alpha}N)=\exp\left(i\sum_{\beta\neq\alpha}^{M}\chi(p_{\alpha},
p_{\beta})\right),\qquad \alpha=1,..M.
\eqno(74)$$
 The energy of corresponding configuration is given by
$$E_{M}=\sum_{\alpha=1}^{M}\sum_{n\neq 0}{{\sinh^2 a}\over{\sinh^2 an}}
(\cos(k_{\alpha}n)-1).$$
For investigating the antiferromagnetic vacuum of the model one should
take $N$ even, $M=N/2$.
Taking logarithms of both sides of (74) and choosing the proper branches, 
one arrives at
$${{Q_{\alpha}}\over N}={{\pi-p_{\alpha}}\over{2\pi}}-
{1\over{\pi N}}\sum_{\beta\neq\alpha}^M \arctan[\varphi(p_{\alpha})
-\varphi(p_{\beta})],$$
where $\{Q\}$ is the set of (half)integers. For antiferromagnetic ground 
state, one assumes as usually that these numbers form uniform string
from $-Q_{max}$ to $Q_{max}$, $Q_{max}=N/4-1/2$ without holes. After introducing
rapidity
variable $\lambda$ by the relation $\lambda=\varphi(k)$ and the function
$\mu(\lambda)$ via the relation $\pi -k=\mu(\lambda)$, the ABA equations
(74) can be written as [38]
$$Q_{\alpha}/N=Z(\lambda_{\alpha}),
\eqno(74a)$$
where
$$Z(\lambda)=(2\pi)^{-1}\mu(\lambda)-{1\over{\pi N}}\sum_{\beta=1}\arctan
(\lambda-\lambda_{\beta}).$$
Following [38], let us go to continuous variable $x=Q_{\alpha}/N$
in the limit $N\to\infty$ and introduce the root density $\sigma_{N}(\lambda)
$ by the relation $\sigma_{N}(\lambda)=dx/d\lambda$.
Differentiating both
sides of (74a) with respect to $\lambda$, one arrives at the following
equation in the limit $N\to\infty$
$$\sigma_{\infty}(\lambda)=(2\pi)^{-1}\mu'(\lambda)-
\int_{-\infty}^{\infty}A(\lambda-\lambda')\sigma_{\infty}(\lambda')d\lambda'
,\eqno(74b)$$
where $A(\lambda)=[\pi(1+\lambda^2)]^{-1}$. The energy per site can be 
written from  as
$$e_{\infty}=\lim_{N\to\infty}N^{-1}E_{N/2}=\int_{-\infty}^{\infty}
\varepsilon(p(\lambda))\sigma_{\infty}(\lambda)d\lambda,
\eqno(75)$$
where
$$\varepsilon(p(\lambda))=2\sinh^2 \kappa\sum_{n=1}^{\infty}
{{\cos np(\lambda)-1}\over{\sinh^2 \kappa n}}.$$
The solution to (74b) can be found via Fourier transform,
$$\sigma_{\infty}(\lambda)=(2\pi)^{-2}\int_{-\infty}^{\infty}
{{e^{i\lambda k}dk}\over{1+e^{-\vert k\vert}}}\int_{-\infty}^{\infty}
\mu'(\tau)e^{-ik\tau}d\tau.$$ 
Substituting it into (75) yields
$$e_{\infty}=(2\pi)^{-2}\int_{-\infty}^{\infty}d\lambda\varepsilon(p(\lambda))
\int_{-\infty}^{\infty}dk{{e^{ik\lambda}}\over {1+e^{-\vert k\vert}}}
\int_{-\infty}^{\infty}\mu'(\tau)e^{-ik\tau}d\tau.$$
Upon choosing variables as $\lambda=\varphi(p),\mu'(\tau)d\tau=-dp'$
and changing the order of integration (it is allowed since the integral
over $\tau$ vanishes sufficiently fast as $\vert k\vert\to\infty$),
 one arrives at the following formula for an energy per site,
$$e_{\infty}=-(2\pi)^{-2}\int_{-\infty}^{\infty}{{dk}\over{1+e^{-\vert k\vert
}}}\int_{0}^{2\pi}dp \varepsilon(p)\varphi'(p)e^{ik\varphi(p)}
\int_{0}^{2\pi}dp'e^{-ik\varphi(p')},\eqno (75a)$$
where the functions $\varepsilon(p)$ and  $\varphi(p)$ are determined as above.
Unfortunately, the integrals in (75b) cannot be evaluated analytically; however,
one can find as $\kappa\to\infty$ that
$$\varphi(p)={1\over 2}\cot{p\over 2} + 2e^{-2\kappa}\sin p+o(e^{-2\kappa}),
$$
$$\varepsilon (p) =2(\cos p-1)+2e^{-2\kappa}(\cos 2p-1)+o(e^{-2\kappa}).$$
Upon substituting these expressions into (75b), the inner integrals are 
calculated analytically up to the order $e^{-2\kappa}$ and final result 
for second-neighbour correlator in the model with nearest-neighbor exchange reads
$$<\vec \sigma_{j}\vec \sigma_{j+2}>=1 -16 \ln 2 +9\zeta(3),$$
where $\zeta $ is the Riemann zeta function which appears in the right-hand
side due to the formula $\int_{0}^{\infty}{{k^2 dk}\over
{1+e^k}}=3/2\zeta(3).$ This result coincides exactly with the expression given by Takahashi [37] who considered the limit of infinite one-site repulsion in the
half-filled Hubbard model.

Another ABA result is the calculation of {\it central charge} $c$ of underlying conformal field model [23]. It is given by the formula for finite-$N$ correction to the energy
of antiferromagnetic ground state
$$\Delta e_{N}=e_{N}-e_{\infty}= -{{\pi c}\over{6N^2}}\xi,
$$
where $\xi$ is the velocity of the lowest-lying elementary excitations. The value of
$\Delta e_{N}$ can be calculated via the equations (74a) where the values of the order
$N^{-2}$ should be carefully taken into account. I would like to mention only the
final result of rather long calculations [23],
$$\Delta e_{N}=-(12N^2)^{-1}\phi_{\infty} +O(N^{-3}),
$$
$$\phi_{\infty}=2\pi i \lim_{\lambda\to\infty}
{{\int_{-\infty}^{\infty}kdk{{\exp(ik\lambda)}\over{1+\exp(-\vert k\vert)}}
\int_{0}^{2\pi}dp\varepsilon(p)\varphi'(p)\exp[-ik\varphi(p)]}
\over
{\int_{-infty}^{\infty}dk{{\exp(ik\lambda)}\over {1+\exp(-\vert k\vert}}
\int_{0}^{2\pi}dp\exp[-ik\varphi(p)]}}.$$
The energy and momentum  of elementary excitations over antiferromagnetic vacuum 
can be also calculated on the base of (74a) under an assumption that this excitation
corresponds to presence of a hole in the sequence of numbers $\{Q\}$. These calculations
result in the formula
$\xi=(2\pi)^{-1}\phi_{\infty}$ which gives the value of the central charge $c=1$ as in
the case of usual nearest-neighbor chain.\\

\noindent {\bf 5. Inhomogeneous lattices.}\\

\noindent It is generally believed that more general dynamical Calogero-Moser 
 systems describing  particles with internal degrees of freedom are integrable.
The motion of particles can be eliminated by arranging them into classical equilibrium
positions. By this way, the first model of {\it inhomogeneous} chain [27] has been obtained
where spin interaction was given by inverse squares of distance between them and
spins were located on equilibrium positions of particles with rational two-body interaction  in the field with a harmonic potential. As for inverse hyperbolic
square exchange, the integrability of the corresponding models is still questionable.
Anyway, there are many indications to this fact as it will be shown later.

The integrability of classical Calogero-Moser systems in some external fields has
been considered in [35]. It was shown there that the Hamiltonians (3) with interaction (5) (with $\kappa=1$ as it can be removed by scaling transformation)  are still integrable if the external field with the potential
$$W(x)=\alpha^2 \cosh(4x)+ 2\beta\cosh (2x)+2\gamma\sinh (2x)
\eqno(76)$$
is added. As for spin chains, the Hamiltonian is still given by
$$H=\sum_{j<k}^{N}h_{jk}P_{jk},
\eqno(77)
$$ 
where $\{P_{jk}\}$ is any representation of the symmetric group $\pi_{N}$, $h_{jk}=
\sinh^{-2}(x_{j}-x_{k})$ and $\{x_{j}\}$ are the coordinates of classical particles 
at equilibrium obeying the equations
$$-2\sum_{k\neq j}h_{jk}c_{jk}+W'(x_{j})=0,
\eqno(78)$$
where
$$c_{jk}=\coth(x_{j}-x_{k}).$$
The first question is to construct the Lax pair for these systems. Consider
the following Ansatz of $(2N\times 2N)$ matrices $(L,M)$ with entries
$$L^{11}=-L^{22}=L_{0}, \quad L^{12}=L^{21}=\psi+\rho, \quad M^{11}=M^{22}=M_{0}+m,
\quad M^{12}=M^{21}=\phi,$$
 where $L_{0}$  and $M_{0}$ is the standard Lax pair for the systems without external
field,
$$(L_{0})_{jk}=(1-\delta_{jk})c_{jk}P_{jk}, \quad (M_{0})_{jk}=(1-\delta_{jk})h_{jk}
P_{jk}-\delta_{jk}\sum_{s\neq j}^{N}h_{js}P_{js}$$
and $\psi,\phi,\rho$ and $m$ are $(N\times N)$ matrices with entries
$$(\psi)_{jk}=\xi(z_{j})\delta_{jk}, \quad \phi_{jk}=\varphi(z_{j})\delta_{jk},\quad
(m)_{jk}=\mu(z_{j})\delta_{jk}, \quad (\rho)_{jk}=(1-\delta_{jk})P_{jk},$$
where $z_{j}=\exp(2x_{j})$. The Lax relation $[H,L]=[L,M]$ is equivalent to the set
of functional equations
$$c_{jk}[\mu(z_{j})-\mu(z_{k})]+[\varphi(z_{j})+\varphi(z_{k})]=0,$$
$$c_{jk}[\varphi(z_{j})+\varphi(z_{k})]+h_{jk}[\xi(z_{j})-\xi(z_{k})]+
\mu(z_{j})-\mu(z_{k})=0.$$
The general solution to this set is given in [36],
$$\mu(z)=\mu_{1}z+\mu_{2}z^{-1},\quad \varphi(z)=-\mu_{1}z+\mu_{2}z^{-1},
\quad \xi(z)=\mu_{1}z +\mu_{2}z^{-1}+\gamma.$$
The potential of an external field reads
$$W(z) =2[\mu_{1}^2 z^2+\mu_{2}^2 z^{-2} +(2\gamma-1/2)(\mu_{1}(z)+\mu_{2}z^{-1})].$$
It contains three free parameters as (76). For the special case of the external Morse
potential  $(\mu_{2}=0)$ the matrix $M$ obeys also the condition $\sum_{j=1}^{2N}
M_{jk}=0$, which guarantees that the integrals of motion can be constructed as
$\{\sum _{j,k}^{2N}(L^{n})_{jk}\}$. In other cases, the existence of the Lax pair
does not imply integrability immediately.

The extra integrals of motion should be some polynomials in the permutations as it
takes place for usual lattice spin models [32]. It turns out that minimal degree of
this polynomial is now equals 3 and the operator
$$I=\sum_{j\neq k\neq l\neq m}^{N}c_{jk}c_{kl}P_{jk}P_{kl}P_{lm}
-{1\over 2}\sum_{j\neq k\neq l}(c_{jl}-c_{kl})^2 P_{jk}+\sum_{j\neq k}^{N}
[F(x_{j})+F(x_{k})]P_{jk}$$
commutes with $H$ if $F$  is a solution of functional equation
$$g(x_{j},x_{k})+g(x_{k}, x_{l})+ g(x_{l}, x_{j})=0,$$
where
$$g(x_{j}, x_{k})=2h_{jk}(F(x_{j})-F(x_{k}))+c_{jk}(W'(x_{j})+W'(x_{k})).$$
The solution is given by the relation $g_(x_{j}, x_{k})=G(x_{j})-G(x_{k})$
and functional equation for the potential
$$c_{jk}(W'(x_{j})+W'(x_{k}))-2h_{jk}(W(x_{j})-W(x_{k}))=G(x_{j})-G(x_{k}).$$
Its general solution just gives the form (76) which supports the hypothesis of
complete integrability of this class of models.

To construct the explicit eigenvalues of the corresponding spin Hamiltonians,
one needs more knowledge about the solutions to equilibrium equations (78). It
can be easily done for special case of the Morse potential
$W(x)=2\tau^2(\exp(4x)-2\exp(2x))$, where these equations have the form [26]
$$-\sum_{k\neq j}^{N}{{z_{k}(z_{j}+z_{k})}\over{(z_{j}-z_{k})^3}}
+\tau^2(z_{j}-1)=0,
\eqno(78a)$$
where the variable $z=\exp(2x)$ is introduced. Following the observation of Calogero
[28], one can assume that the roots $\{z_{j}\}$ of (78a) are given by roots
of some polynomial $p_{N}(z)=\prod_{j=1}^{N}(z-z_{j})$ obeying the second-order
differential equation. In the case of the Morse potential, this equation reads
$$y{{d^2 p_{N}(y)}\over {dy^2}}+(-y+\Gamma +1){{dp_{N}(y)}\over {dy}}+Np_{N}(y)=0,
\quad y=2\tau z,$$
where $\Gamma= 2(\tau-N)+1$. It means that $p_{N}$ are the well-known Laguerre polynomials
$L^{(\Gamma)}_{N}(2\tau z)$. The following properties of their roots will be used:\\
(i) For $\Gamma>-1$, all roots of $L$ are real positive numbers.\\
(ii) As $\Gamma=-N+\varepsilon, \varepsilon\to 0$, all the roots of $L$ approach $0$
with the asymptotic behavior
$$ z_{j}\sim {\rm const}\vert\varepsilon\vert^{1/N}\exp\left({{2\pi ij}\over N}\right).
$$
The rational Calogero spin chain with inverse square exchange [27] is obtained as a limit
of $\tau\to\infty, z_{j}=1+\tau^{-1/2}\xi_{j}$. The lattice points in this limit
are the roots of the Hermite potential $H_{N}(\xi)$. As $\Gamma\to N$, the lattice
becomes equidistant in angles and the model upon rescaling is just the trigonometric
Haldane-Shastry model [26]. Hence the inhomogeneous model defined by the lattice (78a)
can be considered as interpolating between Haldane-Shastry and Polychronakos model.

If one chooses as $\{P_{jk}\}$ in (77) the spin representation of the permutation
group, $P_{jk}=(1+\vec\sigma_{j}\vec\sigma_{k})/2$, the eigenvectors can be treated
as in Sections 3-4. Namely, one can start from the ferromagnetic vacuum $\vert 0>$ with
all spins up and consider the states with given number of down spins $M$,
$$\vert \psi^{(M)}>=\sum_{n_{1}\neq n_{2}...\neq n_{M}}^{N}\psi(n_{1},...n_{M})
\prod_{s=1}^{M}\sigma_{s}^{-}\vert 0>.$$
With the use of the properties of the Laguerre polynomials, one finds that in one-magnon
sector the wave functions can be represented as
$$\psi_{m}(n)\propto z_{n}^{m}{{ L_{N-m-1}^{(\Gamma+2m)}(2\tau z_{n})}\over
{ L_{N-1}^{(\Gamma)}(2\tau z_{n})}}, \qquad m=0,...N-1.$$
The corresponding energies up to universal constant $C_{N}=N(N-1)(3\Gamma +2N-1)/24 $ are given by
$$E_{m}^{(1)}=\epsilon_{m}=-{m\over 2}(\Gamma +m).$$
The two-magnon wave functions can be found analytically and the complete set of 
$N(N-1)/2$ eigenvalues can be written as
$$E_{m,n}^{(2)}=\epsilon_{m}+\epsilon_{n}(1-\delta_{m,n-1}), 0\leq  m<n\leq N-1.$$
In the $M$-magnon sector one can find analytically only some eigenstates
within the Ansatz
$$\psi(n_{1},...n_{M})={{\prod_{\lambda>\mu}^{M}(z_{n_{\lambda}}-z_{n_{\mu}})^2}\over
{\prod_{\nu=1}^{M}p'_{N}(z_{n_{\nu}})}} F(z_{n_{1}},...z_{n_{M}}),$$
where $F$ is some symmetric polynomial in $\{z\}$. It comprises $(N-M+1)![M!(N-2M+1)!]^
{-1}$ eigenvalues which are still additive,
$$E_{\{m_{k}\}}^{(M)}=\sum_{k=1}^{M}\epsilon_{m_{k}}, \qquad m_{k}<m_{k+1}-1, \quad 
0\leq m_{k} \leq N-1.$$
This formula allows one to make the hypothesis about structure of the whole set
of eigenvalues which are described by
$$E_{l_{1}...l_{k}}=\sum_{k=1}^{N-1}\epsilon_{k}l_{k+1}(1-l_{k}),$$
where $\epsilon_{k}=-k(\Gamma+k)/2$ and $\{l_{k}\}=0,1.$ As a consequence of this hypothesis, the 
Hamiltonian $H=2\sum_{j<k}^{N}h_{jk}\vec \sigma_{j}\vec\sigma_{k}$ is unitary 
equivalent to the Hamiltonian of the classical one-dimensional Ising model with
non-uniform magnetic field,
$$H_{I}=\epsilon_{N-1}\sigma_{N}+\sum_{k=0}^{N-2}[\sigma_{k+1}(\epsilon_{k}-
\epsilon_{k+1})-\sigma_{k+1}\sigma_{k+2}\epsilon_{k+1}
\eqno(79)$$
with $\{\sigma_{k}\}=\pm 1.$
This result comprises two above analytical formulae for the spectrum as well as
Haldane-Shastry and harmonic limits and is confirmed by numerical diagonalization
of small lattices up to $N=12$ with several values of the parameter $\tau$.

The simplicity of the spectrum (79) allows one to compute the free energy $f$ as a 
function of the inverse temperature $\beta$ in
the thermodynamic limit upon rescaling the magnon energies with a factor $N^{-2}$ [26].
With the use of quasiparticle dispersion law $\epsilon (x)=-x(\gamma +x)/2$ where
$\gamma=\Gamma/N$ one obtains 
$$f=-{1\over \beta}\left(\int_{0}^{-\gamma}dx \log[1+\exp (\beta\epsilon(x))]
+\int_{-\gamma}^{1}dx \log[1+\exp(-\beta\epsilon(x))]\right),$$
which gives at $\gamma=-1$ the result exactly coinciding with the free energy of
the trigonometric Haldane-Shastry model. 

Coming back to the general potential of an external field (76), one has to start
with the equilibrium equations
$$-\sum_{k\neq j}^{N}{{z_{k}(z_{j}+z_{k})}\over {z_{j}-z_{k}^{3}}}+{1\over 4}\sum_{j=1}^{N}[\alpha^2(z_{j}-z_{j}^{-3})+\beta +\gamma-(\beta-\gamma)z_{j}^{-2}]=0.
\eqno(78b)$$
As in the case of the Morse potential described above, let us introduce the polynomial
$$p_{N}(z)=\prod_{j=1}^{N}(z-z_{j})$$
with the use of the solutions to (78b) and try to identify the differential equation 
to which this polynomial might satisfy. To do this, note that the function $F_{j}(z)$=
$z(z+z_{j})(z-z_{j})^{-3}d\log p_{N}(z)/dz$ has simple poles at $z=z_{k}$ with proper
residues, and the equilibrium equations can be recast in the form
$${\rm res}F_{j}(z)\vert_{z=z_{j}}=2a_{1j}+z_{j}(4a_{2j}-3 a_{1j}^2) +z_{j}^2
(a_{3j}+a_{1j}^{3}-2a_{1j}a_{2j})
\eqno (78c)$$
$$=\alpha^2(z_{j}-z_{j}^{-3})+\beta+\gamma-(\beta-\gamma)z_{j}^{-2},$$
where $a_{\lambda j}=[p'_{N}(z_{j}]^{-1}(d/dz)^{\lambda +1}p_{N}(z)\vert_{z=z_{j}}$.
If one supposes that $p_{N}(z)$ obeys the second-order differential equation
$$z^2 p''_{N}(z)+ w_{1}(z)p'_{N}(z)+w_{2}(z)p_{N}(z)=0
\eqno (80)$$
with some polynomials $w_{1,2}(z)$, one finds upon consecutive differentiations of (80)
with the use of the formula $p_{N}(z_{j})=0$ that the equilibrium equations in the form
(78c) are equivalent to
$$ {d\over dz}[w_{2}+{1/\over 4}(\alpha^2(z^2 + z^{-2})-z^{-2}w_{1}^2)+{1\over 2}
w_{1}'+2(z(\beta+\gamma)+(\beta-\gamma)z^{-1})]=0.$$
This condition is satisfied by $w_{1}(z)=-\alpha(z^2 -1)+(4\alpha^{-1}\beta-\gamma_{1})
z$, $w_{2}(z)=(\alpha-4\beta)z +e_{N}$, where $\gamma_{1}=4\alpha^{-1}\gamma$ and 
parameter $e_{N}$ is still unknown. One of the solutions to (80) is a polynomial of
the degree $N$ if the parameters $\alpha$ and $\beta$ are restricted by
$$\beta=-{{N-1}\over 4}\alpha.$$
The equation (80) is now written as
$$z^2p_{N}''(z)-[\alpha(z^2 -1)+(\gamma_{1}+N-1)z]p'_{N}(z)+(\alpha Nz+e_{N})=0.
\eqno(80a)$$
The substitution $p_{N}(z)=z^{N}+\sum_{l=0}^{N-1}d_{l}z^{l}$ results in the recurrence
relation for $d$- coefficients in the form
$$\alpha d_{l-1}(N-l+1)+d_{l}[e_{N}+l(l-\gamma_{1}-N)]+\alpha(l+1)d_{l+1}=0, \quad l=0,..N.$$
It should be solved under the boundary conditions
$$d_{1}=0,\quad d_{N}=1, \quad d_{N+1}=0.$$
The last condition results in $N$th order equation for the parameter $e_{N}$. The solution must be chosen so as to have all the roots of $p_{N}(z)$ positive. It is unique
since the system of particles which repel each other has only one equlibrium point
being confined in the field with potential (76).

Due to (80a), various symmetric combinations of the roots of (78b) can be expressed
analytically in terms of $\alpha,\gamma_{1}$ and $e_{N}$. In particular, the energy of 
classical equilibrium configuration does not depend on $e_{N}$ and is given by
$$E_{cl}=-{N\over 2}\left({{N^2-1}\over 3}+\gamma_{1}^{2}-2\alpha^2\right).$$
As for corresponding spin chain with the Hamiltonian $H=\sum_{j<k}h_{jk}(\vec \sigma_{j}
\vec\sigma_{k}-1)$, the strategy for finding eigenvalues is the same as for the 
Morse potential described above. However, the information which could be obtained by
this way is much more scarce. In $M$-magnon sectors with $M\leq N/2$, one can use
the ansatz
$$\psi(n_{1},...n_{M})={{\prod_{\lambda>\mu}^{M}(z_{n_{\lambda}}-z_{n_{\mu}})^2}
\over {\prod_{\mu=1}^{M}p'_{N}(z_{n_{\mu}})}}Q(z_{n_{1}},...z_{n_{M}})$$
for multimagnon wave function, and show that the eigenequation can be cast in the
form 
$$\sum_{j=1}^{M}\left\{
z_{j}^2 {{\partial^2}\over {\partial z_{j}^2}}-[\alpha(z_{j}^2 -1)+(\gamma_{1}+N
-3)z_{j}{{\partial}\over{\partial z_{j}}}+2\sum_{j\neq k}^{M}
{{z_{j}^2 \partial/\partial z_{j}-z_{k}^2 \partial/\partial z_{k}}\over
{z_{j}-z_{k}}}\right.$$
$$+\left. M[(M-1)(4M+1)/3-M(\gamma_{1}+N-1)+e_{N}]+\alpha(N-2M)\sum_{k=1}^{M}z_{k}-
2E_{M}\right\}Q=0.$$
For even $N$, the solution at $M=N/2$ $(S_{z}=0)$ is given by $Q$=constant and the
corresponding eigenenergy reads
$$E_{N/2}=1/2\{M[(M-1)(4M+1)/3-M(\gamma_{1}+N-1)+e_{N}]\}.$$
It was verified numerically that for small lattices $(N\leq 8)$ at various sets
of parameters $\alpha$ and $\gamma_{1}$ this is the exact ground state of the antiferromagnetic chain (77). Unfortunately, this approach does not allow to identify
other states and write down such a simple formula for the whole spectrum as in the case
of the Morse potential.\\

\noindent {\bf 6. The related Hubbard chains: are they integrable?}\\

\noindent There are another many-body systems on a lattice  connected
to the Heisenberg-van Vleck spin chains discussed above: the itinerant fermions
of spin 1/2 which interact being at the same lattice site. The corresponding models
are Hubbard chains with the Hamiltonian
$$H_{Hub}=\sum_{j\neq k, \sigma}^{N}t_{jk}c^{+}_{j\sigma}c_{k\sigma}
+2U\sum_{j}^{N}(c^{+}_{j\uparrow}c_{j\uparrow}-1/2)(c^{+}_{j\downarrow}c_{j\downarrow}-
1/2),
\eqno(81)$$
where the operators $c^{+}_{j\sigma}$ create fermion with spin $\sigma$ on the site $j$,
$$ \{c^{+}_{j\sigma}, c_{k\sigma'}\}=\delta_{jk}\delta_{\sigma \sigma'}, \qquad
   \{c_{j\sigma}, c_{k\sigma'}\}=0,
\eqno(82)$$
$t_{jk}\equiv t(j-k)$ is the hopping matrix comprising probability amplitudes for hopping
between sites $j$ and $k$ (it is supposed to be Hermitian) and $U>0$ is the strength of on-site repulsion.

This model was originally introduced by J.Hubbard [42] in three dimensions to describe a metal-insulator transition for systems of fermions with spin. It was found that 1D version (81) is solvable by the Bethe Ansatz [43] in the case of nearest-neighbor hopping
under periodic boundary conditions,
$$t(j)=\delta_{\vert j\vert,1}+\delta_{\vert j\vert, N-1}.
\eqno(83)$$.
The proof of integrability of (81) with the hopping (83), i.e. constructing of the
nontrivial integrals of motion which commute with (81), came much later [44]. There
are two trivial invariants: total number of fermions $M$ and number of fermions
of up (down) spins which are conserved due to $su(2)$ invariance of (81).

The connection with Heisenberg-van Vleck chains discussed above comes in the limit
of infinite $U$ at $M=N$ (half-filled band). In this limit, fermions are not allowed
to occupy the site twice and hop,  i.e. they can interact only via spin exchange. The spin Hamiltonian, which arises in the lowest order in $t/U$, has the form
$$H_{spin}=\sum_{j\neq k}^{N}\vert t_{jk}\vert^2 \vec \sigma_{j}\vec\sigma_{k}.
\eqno(84)$$
It is this relation on which Gebhard and  Ruckenstein (GR) [45] proposed the solvable
model with hopping 
$$t(j)={N\over \pi}{1\over {\sin\left({{\pi j}\over N}\right)}}.
\eqno(85)$$
They were able to guess the simple effective Hamiltonian which comprises all the spectrum
of $H_{Hub}$ with hopping (85) but failed in proving this result analytically. Note
that till now this proof is lacking despite the physical consequences of the GR
hypothesis were investigated thoroughly [46] and numerical calculations also 
support it. Moreover, on the base of (84) yet another model has been proposed [47] with
short-range hopping on the infinite lattice,
$$t(j)=-i\sinh\kappa/\sinh(\kappa j).
\eqno (86)$$
The authors of [47] used the hypothesis of the asymptoic Bethe ansatz for the model (86)
without any proof of integrability and found quite satisfactory properties in the
thermodynamic limit. They showed also that (86) includes, as a limit of $\kappa\to \infty$, the nearest-neighbor hopping (83) on the infinite lattice.

On the base of correspondence with $H_{Hub}$ and its limit (84) one can guess
also the integrability of elliptic model with hopping being some "square root"
of elliptic exchange (12). But in all these cases, one has to find conserved
quantities so as to prove integrability without appeal to any limit or numerical
calculations. This problem is not solved completely till now. But some explicit indications to the integrability are found and will be discussed later.

 In the spectrum of the model with long-range hopping (85), some degeneracies 
were found similar to the degeneracies for the Haldane-Shastry model [48]. This
shows that the model might have additional symmetry besides usual one. For
Haldane-Shastry model, it was found that this symmetry is given by infinite 
vector algebra, the $sl(2)$ Yangian discovered before in [49]. It is natural to
try to find at first the source of degeneracies for the Gebhard-Ruckenstein model (85).
Due to explicit $sl(2)$-invariance of the Hubbard Hamiltonian, it is useful to introduce,
instead of fermion $c$-operators, their bilinear spin-like combinations extending the
concept of spin to different sites. Namely, the product of operators $c^{+}_{j\sigma}
c_{k\tau}$ can be arranged as $2\times 2$ matrix  $(S_{jk})^{\sigma}_{\tau}$ labeled by 
spin indices, which allows one to define the $S$-operators as
$$ S_{jk}^{\alpha}={\rm tr} (\sigma^{*\alpha}S_{jk}), \quad S^{0}_{jk}={\rm tr}(S_{jk}), \quad
S_{j}^{\alpha}=S^{\alpha}_{jj},\quad S^{0}_{j}=S^{0}_{jj},$$
where $\sigma_{\alpha}$ are the Pauli matrices. Note that $S_{j}^{\alpha}/2$ and
$S^{0}_{j}$ are the spin density and fermion density operators. The commutators of these
$S$-operators are
$$[S^{0}_{jk},S^{0}_{lm}]=\delta_{kl}S^{0}_{jm}-\delta_{mj}S^{0}_{lk},
$$
$$[S^{0}_{jk}, S^{\alpha}_{lm}=\delta_{kl}S^{\alpha}_{jm}-\delta_{mj}S^{\alpha}_{lk},
\eqno (87)$$
$$[S_{jk}^{\alpha}, S_{jk}^{\beta}]\delta^{\alpha \beta} \left(
         \delta_{kl} S_{jm}^0 - \delta_{mj} S_{lk}^0 \right)
         + i\varepsilon^{\alpha \beta \gamma} \left(
         \delta_{kl} S_{jm}^\gamma + \delta_{mj} S_{lk}^\gamma \right).
$$ 
There are a lot of other relations between these operators due to their composite nature.
Some of them can be written down explicitly,
$$S_{jk}^{\alpha}S_{lm}^{\alpha}
+  S_{jk}^0 S_{lm}^0 + 2 S_{jm}^0 S_{lk}^0  = 
         4 \delta_{kl} S_{jm}^0 + 2 \delta_{lm} S_{jk}^0,$$
 $$  
     S_{jk}^0 S_{lm}^\alpha + S_{lm}^0 S_{jk}^\alpha +
         S_{lk}^0 S_{jm}^\alpha + S_{jm}^0 S_{lk}^\alpha  = 
         \delta_{jk} S_{lm}^\alpha + \delta_{lm} S_{jk}^\alpha +
         \delta_{lk} S_{jm}^\alpha + \delta_{jm} S_{lk}^\alpha,$$
  $$   S_{jk}^\alpha S_{lm}^\beta + S_{jk}^\beta S_{lm}^\alpha +
     S_{jm}^\alpha S_{lk}^\beta + S_{jm}^\beta S_{lk}^\alpha  = 
         \delta^{\alpha \beta} \left(
         S_{jm}^0 (2\delta_{lk} - S_{lk}^0) + S_{jm}^\gamma S_{lk}^\gamma
         \right),
\eqno(88)$$ 
  $$   - i \varepsilon^{\alpha \beta \gamma} S_{jk}^\beta S_{lm}^\gamma
         - S_{jm}^0 S_{lk}^\alpha + S_{lk}^0 S_{jm}^\alpha  = 
         2 \delta_{lk} S_{jm}^\alpha + \delta_{jk} S_{lm}^\alpha -
         \delta_{lm} S_{jk}^\alpha.$$
These basic relations contain also a whole list of others which appear upon equating
all possible combinations of site indices. In terms of $S$-operators, the Hubbard
Hamiltonian reads
$$H_{Hub}=\sum_{j\neq k}t_{jk}S^{0}_{jk}+U\sum_{j}((S^{0}_{j}-1)^2-1/2).
\eqno (81a)$$
The operators of total spin $I^{\alpha}=1/2 \sum_{j}S^{\alpha}_{j}$ commute with (81a),
their $sl_{2}$ commutation relations are obtained from (87) by summation over lattice sites, $[I^{\alpha}, I^{\beta}]=i\varepsilon^{\alpha\beta\gamma}I^{\gamma}$.
Consider now the operator
$$J^\alpha = {1\over 2} \sum_{j\neq k} \left(
             (f_{jk} + h_{jk}(S_j^0 + S_k^0 - 2))S_{jk}^\alpha
             + g_{jk} \varepsilon^{\alpha \beta \gamma} S_j^\beta S_k^\gamma \right)
\eqno (89),
$$
where $f_{jk}\equiv f(j-k)$ etc. and $g$ and $h$ are odd functions. It is possible to show, with the use of (87-88), that $H_{Hub}$ commutes with $J_{\alpha}$ if the following set of functional equations is satisfied
[50],
$$(g_{jl} - g_{kl}) h_{jk}  = {i\over 2} h_{jl} h_{kl},\quad
       j \neq k \neq l \neq j,$$
    $$  \i U f_{jk}/2h_0 + g_{jk} h_{jk}  =  - {i\over 4}
       \sum_l h_{jl} h_{kl}, \quad
        j \neq k,$$ 
     $$\sum_l (f_{jl} h_{kl} - f_{kl} h_{jl})  =  0,$$
      
     $$t_{jk} =  h_0 h_{jk},$$ 
where $h_{0}$ is a free parameter. It turns out that the only solutions to these
equations just give the trigonometric (finite $N$) and hyperbolic (infinite lattice)
forms of hopping (85) and (86)! In the trigonometric case one finds
$$f_{jk} = 0 , \quad g_{jk} = {1\over2}
         \cot (\pi (j - k)/N), \quad
     h_{jk} = i \sin ^{-1} (\pi (j - k)/N),$$ 
whereas in the hyperbolic case
$$
f_{jk} = {{\sinh(\kappa)(j - k)}\over{U \sinh(\kappa(j - k))}}, \quad
     g_{jk} = {1\over 2} \coth(\kappa(j - k)), \quad
     h_{jk} = i \sinh^{-1} (\kappa(j - k)).$$ 
Note that $J^{\alpha}$ does not depend on $U$ in the trigonometric case. It is natural
to ask of which symmetery does this new vector operator correspond. It turns out that
this symmetry is just Yangian $Y(sl_2 )$  as it can be seen from the commutation relations
$$ [I^{\lambda},J^{\mu}]=i\varepsilon_{\lambda\mu\nu}J^{\nu},$$
$$[J^{\alpha}, K^{\beta}]+[J^{\beta},K^{\alpha}]=0,
\eqno(90)$$
where
$$K^{\alpha}=i\varepsilon^{\alpha\beta\gamma}[J^{\beta}, J^{\gamma}]-4\delta
(I^{\beta})^2 I^{\alpha}$$
and $\delta$=-1 in the trigonometric case and 1 in hyperbolic one. The equation (90) is
just the defining relation for $sl_2$ Yangian. Note also that for all odd functions $t(j)$ there is a canonical transformation 
$$c_{j\downarrow}\to c_{j\downarrow},\quad c_{j\uparrow}\to c_{j\uparrow}^{+},
\quad U\to -U,
\eqno (91)$$
which leaves the Hamiltonian invariant but transforms the Yangian generators $I^{\alpha},
J^{\beta}$ into an independent set of generators $I'^{\alpha},J'^{\beta}$ of another
representation of $sl_{2}$ Yangian. It turns out that these two representations commute
and can be combined to a $Y(sl_{2})\oplus Y(sl_{2})$ double Yangian. The fact of this
commutativity is nontrivial and is of dynamical origin. To verify it and (90), one needs
the explicit form of the functions $f,g,h$ in (89).

The Yangian operator of the nearest-neighbor chain on an infinite lattice found in
[51] can be obtained as a limit of the operator (89) as $\kappa\to\infty$. In the limit
of $U\to \infty$ for half-filled band, where number of fermions coincides with the 
number of lattice sites, one can set $S^{0}_{j}=1$ and recover in the trigonometric
case the Yangian for the Haldane-Shastry model [48]. Thus such rather unlike models
as Haldane-Shastry chain and the infinite Hubbard chain with nearest-neighbor hopping are in fact connected: they could be considered as limiting cases of more general model
with the hopping given by elliptic functions.

It is worth noting that the presence of the Yangian symmetry does not imply integrability. To prove integrability, one has to construct the set of {\it scalar}
currents with number of its elements at least equal to the number of lattice sites.
It was proved for the Hubbard model with nearest-neighbor hopping by finding
its connection to spin ladder and with two coupled six-vertex models [44]. These
methods definitely do not work for the Gebhard-Ruckenstein model and its hyperbolic
counterpart. One has to find another method for constucting integrals of motion.

To provide examples of the conserved currents which might exist for 
some choice of the hopping matrix,
 consider the ansatz
$$
J=\sum_{j\neq k}^{N}[A_{jk}S_{jk}^{0}+B_{jk}(S_{j}^{0}S_{k}^{0}-\vec S_{j}
\vec S_{k})+D_{jk}(S_{j}+S_{k}^{0})S_{jk}^{0}+E_{jk}(S_{jk}^{0})^2].
\eqno(92)$$
which is most general scalar operator bilinear in $\{S\}$. By definition,
$A_{jk}\equiv A(j-k)$ etc. The condition $[H_{Hub},J]=0$ with the use of (87-88)
can be cast into the form of two functional equations
$$
4t_{jk}(B_{lk}-B_{jl})+(t_{jl}D_{lk}-D_{jl}t_{lk})=0,
\eqno(93)$$
$$
2(t_{jk}E_{kl}+t_{kj}E_{jl})+(t_{jl}D_{kl}+t_{kl}D_{jl})=0,
\eqno(94)$$
the definition of $A$
$$
A_{jk}=-2D_{jk}+(2U)^{-1}[-8t_{jk}B_{jk}+2t_{kj}E_{jk}-r_{jk}],
$$
where 
$$ r_{jk}=\sum_{l\neq j ,k}^{N}
t_{jl}D_{lk},$$
and several "boundary" equations for $t,B$ and $D$:
$$\sum_{l\neq j,k}^{N}(t_{jl}A_{lk}-A_{jl}t_{lk})=0,$$
$$\sum_{k\neq j}^{N}(t_{jk}D_{kj}-D_{jk}t_{kj})=0,$$
$$\sum_{k\neq j}^{N}(t_{jk}A_{kj}-t_{kj}A_{jk})=0.$$
The first functional equation (93) is just the Calogero-Moser functional equation (10)
with known general analytic solution. The second functional equation (94) always
has solutions for $E_{jk}$ if $t$ and $D$ are given by solutions of (93). 
Each function in these and "boundary" equations can be expressed via basic solution to
(93), and the role of "boundary" equations is to specify the real period of the corresponding Weierstrass functions, which turns out to be $N$. The basic solution 
reads
$$\psi(x)={{\sigma_{N}(x+\lambda)}\over{
\sigma_{N}(x)\sigma_{N}(\lambda)}}e^{\nu x}.
\eqno(95)$$
The other functions in (92-93) are expressed as (recall that $t_{jk}\equiv t(j-k)$ etc.)
$$t(x)=t_{0}\psi(x),\qquad B(x)=-{d\over4}\psi(x)\psi(-x),$$
$$D(x)=d[\psi'(x)-({{h\wp'_{N}(\lambda)}\over2}+\zeta_{N}(\lambda)+\nu)\psi(x)],\quad
E(x)={{d\psi^2(x)}\over2}[1-h\psi(x+\lambda)\psi(-x-\lambda)],$$
$$r(x)=t_{0}d\psi(x)[-(N-3)\wp_{N}(x)+h_{1}(N-2)\tau(x)+(\tau(x)-h_{1})
(2x\zeta_{N}(N/2)-N\zeta(x))+s],$$
$$\tau(x)=\zeta_{N}(x+\lambda)-\zeta_{N}(x)-\zeta_{N}(\lambda), \quad h_{1}=h\wp_{N}'(\lambda)/2,
\quad s=-(N-2)\wp_{N}(\lambda)-\sum_{l=1}^{N-1}\wp_{N}(l),$$
where $\sigma_{N},\zeta_{N}$ and $\wp_{N}$ are the Weierstrass elliptic functions determined by
the periods $\omega_{1}=N, \omega_{2}=i\pi/\kappa$, $\lambda=i\alpha$ or $i\alpha+N/2$, $\nu=i\beta$,
 $\kappa,d,h,\alpha,\beta$ being arbitrary real parameters. At these conditions,
the hopping matrix is Hermitean. Besides this general solution, there are the degenerate
rational, hyperbolic and trigonometric ones, which correspond to one or two periods of
the Weierstrass functions. In the first two cases, the lattice should be infinite.
Checking the absence of "boundary" terms is nontrivial task with key formula
$$[\wp(y+\lambda)-\wp(\lambda)][\zeta(x-y)-\zeta(x+\lambda)+\zeta(y)+\zeta(\lambda)]+$$
$$[\wp(x+\lambda)-\wp(\lambda)][\zeta(y-x)-\zeta(y+\lambda)+\zeta(x)+\zeta(\lambda)]=
\wp'(\lambda).$$
These formulas for $t,B,D,E,r,A$ define the scalar current (92) for the model with
elliptic hopping which comprises all the hopping matrices (83),(85,86) considered above.
At $\lambda$ being the half-period of the Weierstrass $\wp_{N}$ function, the function
$\psi(x)$ becomes odd and yet another current is obtained from (92) by the canonical
transformation (91).

The presence of scalar currents commuting with Hamiltonian is the first evidence of
the integrability for the Hubbard models with the hopping (95) presenting the
"square root" for the elliptic exchange in Heisenberg-van Vleck chains. It is possible
to find the corresponding two-fermion function analytically [52]. However, it is seen
also that the construction of higher scalar currents is extremely hard problem and
many-fermion wave functions should be also cumbersome and complicated. Till now, 
nothing is known even about ground-state wave function of the simplest trigonometric
Gebhard-Ruckenstein model: it is neither of Jastrow-type as for the Haldane-Shastry
model nor of Bethe ansatz form as in the case of the hopping (83).\\

\noindent {\bf 7. Concluding remarks.}\\

\noindent The main known facts about the integrable Heisenberg-Van Vleck chains with variable range exchange and related Hubbard models were reviewed. Many questions in their theory are still open.

As concerns the integrability of these models, understanding  it from the Yang-Baxter
viewpoint is highly desirable. For the spin chains, it is quite probably that the corresponding $R$ matrix is the same as in [33]. The problem of mutual commutativity
of the set of operators (18) might be solved in this way. Nothing is known for the integrability of elliptic Hubbard chains except of the simplest conserved current (92) and two-fermion wave function. 

The model with hyperbolic exchange on infinite line should have rich variety of
multimagnon bound states which are given by solutions to transcendental equations
$1-(2\kappa)^{-1}[f(p_{j})-f(p_{k})]=0$ as it follows from (55). It would be of interest
to find more simple way of constructing eigenfunctions of the Calogero-Moser Hamiltonian
with inverse square hypebolic particle interaction.

The exact equations of Bethe ansatz type for the case of periodic boundary conditions
are too complicated at the present stage of finding solutions to quantum elliptic
Calogero-Moser equation at $l=1$. One cannot exclude the possibility of discovering
their more simple form which would be of use to verify the hypothesis of asymptotic
Bethe ansatz in the thermodynamic limit. The construction described above does not
allow neither to do that nor to establish the correspondence with the trigonometric
Haldane-Shastry model. 

In the models on inhomogeneous lattices, the main problem also consists in finding
the proof of integrability for the most general potential of the external field (76).
The simple formula for the spectrum for the case of the Morse potential, which comprises
rational and Haldane-Shastry models, still waits analytical confirmation. If one
would find the explicit form of the unitary transformation of the basic Hamiltonian
to its simple effective form (79), a lot of results about various correlation functions
would be obtained for the Haldane-Shastry chain.

The only known results about the spectra of the related Hubbard models are given by
original work of Gebhard and Ruckenstein [45]. The trigonometric and hyperbolic versions
both have the $sl_2 \otimes sl_2$ Yangian symmetry and scalar integrals of motion (92).
The most challenging problem is to prove the integrability and find the Bethe-ansatz-
like formulas for the spectrum of the most general Hubbard model with elliptic
hopping (95). Its solution could clarify the algebraic nature of the integrability
of all the models under present discussion. \\

\noindent {\bf Acknowledgments}\\

\noindent
I would like to thank  Jaroslav Dittrich, Bernd D\"orfel, Holger
Frahm, Frank G\"ohmann and Ryu Sasaki for many fruitful discussions and collaboration.
The work has been supported by the Japan Society for Promotion of Science.\\
 
\noindent {\bf References} 
\begin{enumerate}
\item
W.Heisenberg. Z.Phys. 49,619 (1928)
\item
P.A.M.Dirac. Proc.Roy.Soc. 123A, 714 (1929)
\item
J.H. van Vleck. The Theory of Electric and Magnetic Susceptibilities.
Oxford at the Clarendon Press, 1932
\item
H.A.Bethe. Z.Phys. 71,205 (1931)
\item
M.Gaudin. La fonction d'onde de Bethe. Paris: Masson 1983
\item
L.D.Faddeev, in: Recent Advances in Field Theory and Statistical Mechanics,
p.561. Amsterdam: North-Holland 1984
\item
V.E.Korepin, N.M.Bogoliubov and A.G.Izergin. Quantum Inverse Scattering Method
and Correlation Functions. Cambridge University Press, Cambridge 1993
\item
F.Calogero. J.Math.Phys. 12,419 (1971)
\item
B.Sutherland. J.Math.Phys. 12,246,251 (1971); 
Phys.Rev. A5,1372 (1972)
\item
F.Calogero, C.Marchioro and O.Ragnisco. Lett.Nuovo Cim. 13,383 (1975)
\item
F.Calogero. Lett.Nuovo Cim. 13,411 (1975); J.Moser. Adv.Math. 16,197 (1975)
\item
I.M.Krichever. Funct.Anal.Appl. 14,282 (1980)
\item
M.A.Olshanetsky and A.M.Perelomov. Phys.Rep. 94,313 (1983)
\item
F.D.M.Haldane. Phys.Rev.Lett. 60,635 (1988); Phys.Rev.Lett. 66,1529 (1991)
\item 
B.S.Shastry. Phys.Rev.Lett. 60,639 (1988)
\item
F.D.M.Haldane, in the Proceedings of the 16th Taniguchi Symposium on Condedsed Matter
Physics, ed. by A.Okiji and N.Kawakami; cond-mat/9401001
\item
V.I.Inozemtsev. J.Stat.Phys. 59,1143 (1990)
\item
V.I.Inozemtsev and N.G.Inozemtseva. J.Phys. A24,L859 (1991)
\item 
J.Sekiguchi. Publ.RIMS Kyoto Univ. 12,455 (1977)
\item
O.A.Chalykh and A.P.Veselov. Commun.Math.Phys. 126,597 (1990)
\item
 V.I.Inozemtsev. Commun.Math.Phys. 148,359 (1992)
\item
V.I.Inozemtsev. Lett.Math.Phys. 28,281 (1993)
\item
V.I.Inozemtsev and B-D. D\"orfel. J.Phys. A26,L999 (1993)
\item
J.Dittrich and V.I.Inozemtsev. J.Phys. A26,L753 (1993)
\item
E.K.Sklyanin. Progr.Theor.Phys.Suppl. 118,35 (1995)
\item
H.Frahm and V.I.Inozemtsev. J.Phys. A27,L801 (1994)
\item 
A.P.Polychronakos. Phys.Rev.Lett. 70,2329 (1993); H.Frahm. J.Phys. A26,L473
(1993)
\item
F.Calogero. Lett.Nuovo Cim. 19,505 (1977)
\item
V.I. Inozemtsev. Phys.Scripta 39,289 (1989)
\item
V.I.Inozemtsev. J.Phys. A28,L439 (1995)
\item
V.I.Inozemtsev. J.Math.Phys. 37,147 (1996)
\item
V.I.Inozemtsev. Lett.Math.Phys. 36,55 (1996)
\item
K.Hasegawa. Commun.Math.Phys. 187,289 (1997)
\item
J.Dittrich and V.I.Inozemtsev. Mod.Phys.Lett. B11,453 (1997)
\item
V.I.Inozemtsev. Phys.Lett. 98A,316 (1983)
\item
V.I.Inozemtsev and N.G.Inozemtseva. J.Phys. A30,L137 (1997)
\item
M.Takahashi. J.Phys. C10,1289 (1977)
\item
J.Dittrich and V.I.Inozemtsev. J.Phys. A30, L623 (1997)
\item
G.Felder and A.Varchenko. Int.Math.Res.Notices 5,222 (1995)
\item
V.I.Inozemtsev. Regular and Chaotic Dynamics 5,236 (2000); math-ph/9911022
\item
B.Sutherland and B.S.Shastry. Phys.Rev.Lett. 71,5 (1993)
\item 
J.Hubbard. Proc.R.Soc.London A276,238 (1963)
\item
E.H.Lieb and F.Y.Wu. Phys.Rev.Lett. 20,1445 (1968)
\item 
B.S.Shastry. J.Stat.Phys. 50,57 (1988
\item
F.Gebhard and A.E.Ruckenstein. Phys.Rev.Lett. 68,214 (1992)
\item
F.Gebhard, A.Girndt and A.E.Ruckenstein. Phys.Rev. B49,10926 (1994)
\item
P.-A.Bares and F.Gebhard. Europhys.Lett. 29,573 (1995)
\item
F.D.M.Haldane et al. Phys.Rev.Lett. 69,2021 (1992)
\item
V.G. Drinfel'd. Soviet.Math.Dokl. 32,254 (1985)
\item
F.G\"ohmann and V.I.Inozemtsev. Phys.Lett. A214,161 (1996)
\item
D.B.Uglov and V.E.Korepin. Phys.Lett. A190,238 (1994)
\item 
V.I.Inozemtsev and R.Sasaki. Phys.Lett. A289,301 (2001)
\end{enumerate}
\end{document}